\documentclass[%
 reprint,
 superscriptaddress,
 amsmath,amssymb,amsfonts,
 aps,
]{revtex4-2}

\usepackage{kotex}
\usepackage{array, booktabs, tabularx}

\usepackage{siunitx}
\DeclareSIUnit[number-unit-product = {\,}]
\cal{cal}
\DeclareSIUnit\kcal{\kilo\cal}

\newcolumntype{Z}{>{\centering\let\newline\\\arraybackslash\hspace{0pt}}X}
\newcolumntype{Q}{>{\raggedright\let\newline\\\arraybackslash\hspace{0pt}}X}
\newcolumntype{G}{>{\raggedleft\let\newline\\\arraybackslash\hspace{0pt}}X}
\usepackage{graphicx}
\usepackage{dcolumn}
\usepackage{bm}
\usepackage{hyperref}
\hypersetup{
    colorlinks=true,
    linkcolor=blue,
    filecolor=magenta,      
    urlcolor=cyan,
    }
\begin{document}

\title{A2I Transformer: Permutation-equivariant attention network for pairwise and many-body interactions with minimal featurization
}


\author{Ji Woong Yu}
\thanks{These authors contribute equally to this work.}
\affiliation{School of Chemical and Biological Engineering, Institute of Chemical Processes, Seoul National University, Seoul 08826, Korea}
\author{Min Young Ha}
\thanks{These authors contribute equally to this work.}
\affiliation{School of Chemical and Biological Engineering, Institute of Chemical Processes, Seoul National University, Seoul 08826, Korea}
\author{Bumjoon Seo}
\thanks{seo89@purdue.edu; wblee@snu.ac.kr}
\affiliation{Davidson School of Chemical Engineering, Purdue University, West Lafayette, Indiana 47906, United States}
\author{Won Bo Lee}
\thanks{seo89@purdue.edu; wblee@snu.ac.kr}
\affiliation{School of Chemical and Biological Engineering, Institute of Chemical Processes, Seoul National University, Seoul 08826, Korea}

\date{\today}

\begin{abstract}

    The combination of neural network potential (NNP) with molecular simulations plays an important role in an efficient and thorough understanding of a molecular system's potential energy surface (PES). However, grasping the interplay between input features and their local contribution to NNP is growingly evasive due to heavy featurization. In this work, we suggest an end-to-end model which directly predicts per-atom energy from the coordinates of particles, avoiding expert-guided featurization of the network input. Employing self-attention as the main workhorse, our model is intrinsically equivariant under the permutation operation, resulting in the invariance of the total potential energy. We tested our model against several challenges in molecular simulation problems, including periodic boundary condition (PBC), $n$-body interaction, and binary composition. Our model yielded stable predictions in all tested systems with errors significantly smaller than the potential energy fluctuation acquired from molecular dynamics simulations. Thus, our work provides a minimal baseline model that encodes complex interactions in a condensed phase system to facilitate the data-driven analysis of physicochemical systems.

\end{abstract}

\maketitle

\section{\label{sec:introduction} Introduction}

    The fundamental behaviors of a physical system come from how the microscopic components of the system interact among themselves. To perform any molecular simulations, approximations to these interactions and the resulting potential energy surfaces are required where the choice of simulation methodologies is up to the purpose of the simulation and computational cost. For example, the density functional theory (DFT) captures $n$-body interaction to predict chemical properties at the \textit{ab initio} accuracy, but this comes at the price of demanding computational cost, which limits the application to several hundreds of atoms. On the other hand, interatomic potential models with pairwise-additive potential energy functions are used in the molecular simulations of which the parameters are mostly derived by fitting to experimental data such as density, the heat of vaporization, and solvation free energy \cite{jorgensen1988opls,jorgensen1996development,wang2004development,vanommeslaeghe2010charmm}. Compared against DFT, these classical force fields allow extended time and length scales for the simulations at the cost of accuracy. 

    In contrast to the approaches mentioned above in which the description of the interaction is dealt with within the framework of existing experimental or theoretical physics, data-driven approaches are recently gaining significant interest ever since the pioneering high-dimensional neural network potential (HDNNP) by Behler and Parrinello \cite{behler2007generalized}. Input to the HDNNP is an array of localized structure descriptors that encodes the atomic environment in terms of the relative position of an atom to its neighbors. For example, it is common to use displacement vectors to neighboring particles or angles between two bonded neighbors \cite{musil2021physics}. Developments have been made in formulating the ideal descriptors for atomic environments \cite{smith2017ani,zhang2018deep,zhang2018end,yao2018tensormol,bartok2010gaussian} since the key to performance is making a good feature that the model can process into the particle's potential energy.

    Although descriptor-based NNP is widely popular, its success is significantly limited by domain knowledge and trial-and-error since not only the content of the descriptor but also the representation of it matters. For example, naive pairwise distance or angle of a triplet of atoms is not sufficient; the standard approach is to process the distance by Gaussian kernels \cite{behler2007generalized, behler2011atom}. Moreover, even if the information content of the descriptor is complete, it cannot predict well if its representation is not suitable, as in any other descriptor-based applications such as particle classifier \cite{yoon2018probabilistic} or order parameters \cite{schoenholz2016structural,seo2018driving,boattini2020autonomously}. Thus, both the content and representation of descriptors are important for a descriptor-based NNP.
    
    More recently, message passing neural network potential (MPNNP) \cite{gilmer2017neural,schutt2017quantum,schutt2017schnet,lubbers2018hierarchical,zubatyuk2019accurate} has been introduced, which is another kind of NNP that replaces the predefined descriptors with learnable ones by constructing molecular graphs to update atomic feature vectors through interaction blocks iteratively. A key aspect of MPNNP is that it does not appeal to the extensive domain knowledge of experts to make effective descriptors but gives the burden of feature engineering to the neural network \cite{bartok2013representing}. MPNNPs exploit the graph structure to iteratively update the hidden state at each node by incorporating the message function from its neighboring nodes \cite{gilmer2017neural}. Instead of encoding the necessary information into expert-guided features, MPNNPs utilize the communication between the hidden states to learn the relevant representation from less processed input features: hence it is categorized as representation learning.



    In the context of bias-variance trade-off \cite{munson2009feature}, it is generally advised to suppress the variance by reducing the size of the features, up to the point where the increase in bias significantly undermines the prediction power. The key achievement of MPNNP is that it significantly outperformed descriptor-based NNPs \cite{gilmer2017neural} while keeping a substantially smaller set of input features. Another gain of carrying a smaller set of features is that it helps users keep track of the features' contribution to the final prediction, in line with the timely context of explainable AI (XAI).

    Inspired by the above series of discussions, we raise two questions to further improve the advantages of MPNNP in terms of representation learning and XAI. First, what is the limit of feature reduction, e.g., is it possible to avoid calculating the pairwise distance between atoms? Secondly, is it possible to directly obtain per-atom variables as network output, facilitating the detailed analysis of the locally heterogeneous environment or per-atom contributions to the system?
    
     
    \begin{figure}[!htb]
    \includegraphics[width=3.4in, height=3.4in]{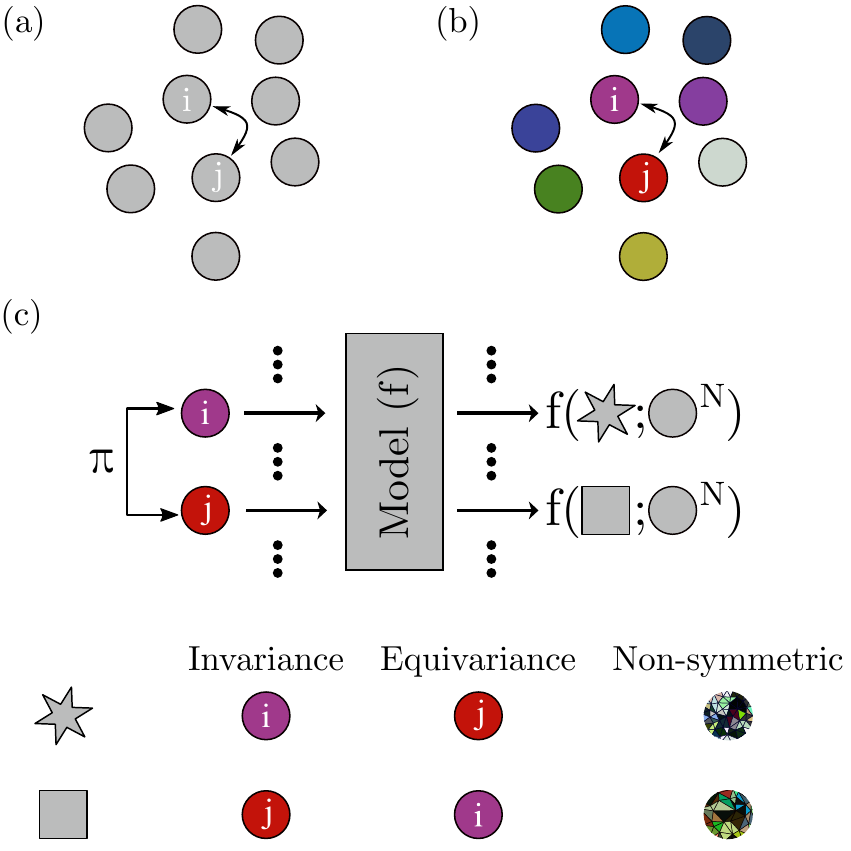}
    \caption{\label{fig:color_particle} Schematic of (a) indistinguishable particles and (b) distinguishable particles. Permutation operation ($\pi$) over particles affect nothing in (a) but changes the configuration in (b). (c) Three types of permutation symmetry upon permutation operation. Given the model function ($f$), $f$ is permutation invariant if the output does not change upon the action of $\pi$ (i.e. $f(x)=f(\pi x)$). $f$ is permutation equivariant if the output also permutate in the same way as input (i.e. $\pi f(x)=f(\pi x)$). The rest of the cases has no relation with permutation symmetry and the outcome is the composite of several influences from the input.}
    \end{figure}

    The above discussion naturally leads us to push the power of representation learning further to develop an end-to-end model that directly predicts the per-atom energy from the raw coordinates of particles. \cite{samek2019towards}. Two levels of physical constraints should be obeyed in this approach: first, the total energy of the system should be invariant under the permutation of equivalent particles; second, the energies of atoms $i$ and $j$ should be swapped under the exchange of atomic labels $i$ and $j$, i.e., equivariant under the permutation. Thus, the input should be treated as a set rather than an ordered sequence. However, simple neural networks such as multilayer perceptron (MLP) mix and entangle the outputs from the previous layer as the input goes deep into the layers; instead of $n$ identical particles (see Fig.~\ref{fig:color_particle}(a)), what the network encounter is a group of $n$ \textit{colored} particles (see Fig.~\ref{fig:color_particle}(b)). In our case, the model should be equivariant (see Fig.~\ref{fig:color_particle}(c)) under symmetric operations, otherwise the complexity of the task increases by $\sim n!$ and makes the learning almost impossible even for a handful of particles.

    \begin{figure*}[!htb]
    \includegraphics[width=6.8in, height=5.0in]{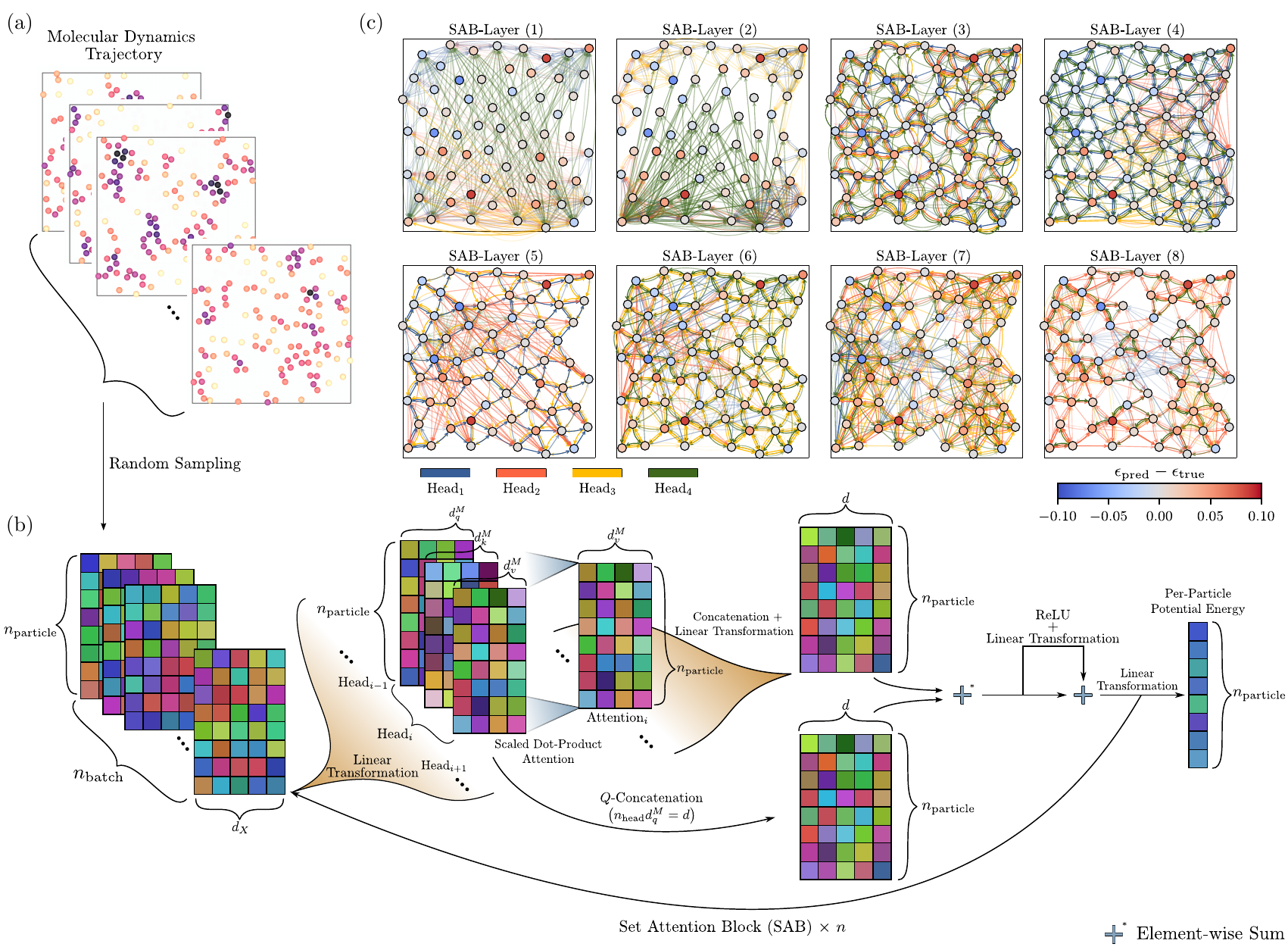}
    \caption{\label{fig:network_description} (a) The prepared molecular dynamics snapshots are randomly sampled to compose the input batch. (b) The schematic diagram of the network architecture, (c) Visualization of attention array ($\bm{Q}\bm{K}^\mathsf{T}$) in 2D fixed boundary LJ ($N=64$) system. The color of the nodes denotes the difference between the true and predicted energy for which the corresponding color bar is shown at the right bottom of (c). Each head shows a different pattern of attention. Note that only a few highest attraction values starting from one node or those with an attraction value higher than $0.08$ are visualized. Also, the transparency of the edge indicates the relative intensity of attention.}
    \end{figure*}

    In this paper, we present an attention-based NNP model with the atom coordinates as the only input. Attention is an importance score of a pair of elements. Since the early success of the attention-based Transformer \cite{vaswani2017attention} in natural language processing, a lot of models derived from the Transformer architecture have achieved state-of-the-art in diverse area including image recognition \cite{dosovitskiy2020image, liu2021swin}, reinforcement learning \cite{chen2021decision}, and protein folding \cite{AlphaFold2021}. We are directly inspired by the permutation-invariant Set Transformer \cite{lee2019set} in which pairwise- or higher-order interaction can be captured by stacking of self-attention units. Moreover, as the name suggests, "attention" can be used as a useful latent space metric sending signals that need to be attended to.
    
    We tested our model in various simulation setups. Starting from a fixed boundary system with simple pair interactions, we resolved various challenges to verify the applicability of our model so that our model satisfies requirements for generic molecular simulations. First, we introduced position encoding to handle periodic boundary condition (PBC), which is widely used to estimate bulk properties from a finite-sized simulation. Secondly, stacking multiple self-attention blocks allowed the model to capture higher-order interactions, as shown in the example of Stillinger-Weber potential. Third, we extended our model to a binary Lennard-Jones fluid system by adding an extra dimension for the particle type, thereby rendering the model to handle multi-component systems with numerous atom types. In all systems, the prediction error was significantly smaller than the intrinsic fluctuation of per-atom energy in the molecular dynamics simulations. Through this series of demonstrations, we stress the versatility of our model.
    
    This paper is organized as follows. We firstly introduce our setup in section \ref{sec:method}. We demonstrate that the ability of our model to accurately predict the potential energy of each particle and its permutation equivariance in a simple toy system in section \ref{sec:toy_model}. To demonstrate the generalizability of our model, we prepare three variants of the simple system, each representing a frequently encountered challenge in molecular simulations in section \ref{sec:extended}. We summarize our work and comment on future outlooks in the section \ref{sec:future}.

\section{\label{sec:method} Method and Preparation}
    
    In this section, we will explain our neural network model and the task.
    
    \subsection{\label{subsec:Set Transformer} Set Attention Block}
    Our model is a modification, and an extension of Set Transformer \cite{lee2019set}, which is a variant of Transformer \cite{vaswani2017attention}, modified to operate on a set that is invariant under permutation of its elements. We recap here some of the details of the model.
    
    Attention is a category of correlation metrics which measures the directed importance between a pair of elements \cite{graves2014neural, bahdanau2014neural,luong2015effective, vaswani2017attention}. Notably, the most widely used attention is \textit{scaled-dot product attention} (Eq.~\eqref{eq:scaled_dot_attention}) \cite{vaswani2017attention}. Given $n$ query vectors in $d_q$-dimension ($\bm{Q} \in \mathbb{R}^{n \times d_q}$), $\bm{Q}\bm{K}^\mathsf{T}$ correlates each query vector with every key vector in $m$ $d_q$-dimension keys ($\bm{K} \in \mathbb{R}^{m\times d_q}$). The resultant attention matrix ($\bm{Q}\bm{K}^\mathsf{T}$) goes through activation function $\omega$ and is weight averaged with $m$ $d_v$-dimension value vectors ($\bm{V}\in\mathbb{R}^{m\times d_v}$). The activation function $\omega$ is $\textrm{softmax}\left(\cdot/\sqrt{d_v}\right)$
    
    \begin{equation}
    \label{eq:scaled_dot_attention}
        \textrm{Att}\left(\bm{Q}, \bm{K}, \bm{V}; \omega\right) = \omega\left(\bm{Q}\bm{K}^\mathsf{T}\right)\bm{V}
    \end{equation}
    
    Transformer \cite{vaswani2017attention} is a multi-head attention model which has more expressive power. In multi-head attention, each dimension is divided into different heads. The heads do not share learnable parameters and are independent of each other.
    
    
    \begin{align}
        \textrm{Multihead}\left(\bm{Q}, \bm{K}, \bm{V}\right)=\textrm{concat}\left(\bm{O}_1, \bm{O}_2, ..., \bm{O}_h \right)\bm{W}^O \label{eq:multihead_attention}\\
        \textrm{where } \bm{O}_i=\textrm{Att}\left(\bm{Q}\bm{W}_i^Q,\bm{K}\bm{W}_i^K,\bm{V}\bm{W}_i^V;\omega_i\right) \label{eq:head}
        \end{align}
    
    Each head has its own attention $\bm{O}_j$. The attentions are assembled into multi-head attention by concatenation and linear transformation with $\bm{W}^O\left(\in \mathbb{R}^{d_v\times d_v}\right)$. Query, key, and value at each head have condensed dimension $d_q^M$, $d_q^M$, and $d_v^M$, respectively, using separate learnable linear transformation ($\bm{W}_i^Q \in \mathbb{R}^{d_v\times d_v^M}$, $\bm{W}_i^Q \in \mathbb{R}^{d_v\times d_v^M}$, $\bm{W}_i^Q \in \mathbb{R}^{d_v\times d_v^M}$). The most common choice for the condensed dimension is dividing original dimension by the number of heads ($h$) (i.e.\ $d_q^M=d_q/h$ and $d_v^M=d_v/h$).
    
    Set Transformer \cite{lee2019set} defines Multi-head Attention Block (MAB) using multi-head attention described above. Given $n$ $d$-dimensional vectors $\bm{X}$ and $\bm{Y}$ $\left(\in\mathbb{R}^{n\times d}\right)$,
    \begin{align}
        &\textrm{MAB}(\bm{X},\bm{Y})=\bm{H}+\textrm{rFF}\left(\bm{H}\right)\label{eq:MAB}\\
        &\textrm{where } \bm{H}=\bm{X}+\textrm{Multihead}\left(\bm{X},\bm{Y},\bm{Y};\omega \right)\label{eq:H}
    \end{align}
    
    where rFF is the row-wise feed-forward layer. The particular case ($\bm{X}{=}\bm{Y}$) of MAB is Set Attention Block (SAB) in which each element, which is $d$-dimensional vector, attends to every element, including itself. Note that the SAB block is permutation equivariant. In Lee et al. \cite{lee2019set}, pooling operation is performed at the end of the neural network layers, and the model gains permutation invariance. In this work, since we wish the model to capture the per-atom potential energy rather than the total potential energy of the system, we deleted the pooling layer to render the model permutation equivariant.
    
    \begin{equation}
        \label{eq:SAB}
        \textrm{SAB}(\bm{X})=\textrm{MAB}(\bm{X},\bm{X})
    \end{equation}
    
    \subsection{Task Description}
    
    We are interested in predicting the energy of each particle from any particle-based simulation without any \textit{a priori} knowledge of the potential (e.g., order of the potential, functional form, etc.). This requires that the input of the model be only the most unprocessed form of data from a particle-based simulation, which is the coordinates of each particle. Thus, we choose our model input, $X$, as an array consisting of $n$ $d$-dimension particle coordinates ($X\in\mathbb{R}^{n\times d}$) where $n$ is the number of particles and $d$ is the dimension of the particle system. Corresponding output which is per-particle energy, $U$, is an array consisting of $n$ one-dimensional scalars ($U\in \mathbb{R}^{n \times 1}$). Thus, our goal is to make a multiobjective model, $f$
    
    \begin{equation}
    \label{eq:model}
        f:X(\in\mathbb{R}^{n\times d}) \rightarrow U(\in\mathbb{R}^{n \times 1})
    \end{equation}
    
    \subsection{Model, Data Preparation, and Learning}
    
    The primary target property for Set Transformer \cite{lee2019set} was the permutation invariance, whereas our task requires the permutation-equivariant property. In the original Set Transformer, there exists a pooling layer that turns the permutation-equivariant network into the permutation-invariant one. By only taking out the pooling layer, we can obtain a permutation-equivariant network. 
    We stacked a total of $8$ SAB layers in all cases and omitted the \textit{layer normalization} for the simplicity of the model in Eq.~\eqref{eq:MAB} and~\eqref{eq:H}. At the end of the last SAB layer, the linear transformation was performed to get the desired dimension of $U$. The output from each SAB was set to have a hidden dimension $d_v=32$ unless noted otherwise. The number of heads was set to $h=4$. We masked the attention matrix by making the diagonal elements to zero to reflect the fact that the interaction is between particles. See FIG.~\ref{fig:network_description}(b) for the schematic model architecture.
    
    For all systems, we run molecular dynamics simulation with either LAMMPS \cite{plimpton1995fast} or HOOMD \cite{anderson2020hoomd}. The system-specific details will be provided in the following sections. For training the model, over $2$ million snapshots were prepared. For the test, $25{,}600$ snapshots were prepared. The model was trained for $100{,}000$ batches with batch size $128$. Snapshots are randomly chosen without checking that the snapshot was already used for another batch (FIG.~\ref{fig:network_description}(a)). Note that our model can be trained with snapshots generated in other methods. For example, snapshots generated by Monte Carlo simulation or even randomly seeded particles can be used as long as the potential energy can be estimated and the sampling quality is ensured. Also, the snapshots need not be sequential in time, nor need be prepared in chronological order. For each batch, we performed optimization repeatedly from $10$ to $100$ times before moving on to another batch to reduce data transfer between storage and GPU. This might cause a well-known overfitting issue, but we did not suffer from it with the above-mentioned repeats.
    
    Adam optimizer \cite{kingma2014adam} is used for all cases with the learning rate of $0.0002-0.0005$. Four GPU devices (NVIDIA GTX 1080 Ti, RTX 2080 Ti, RTX 3090, and Tesla A100) were used for the training and the test. We confirmed that the results from all devices agree. All numeric operations were performed in single precision.
    
    \subsection{Metric}
    
    We use four metrics to quantify and compare the performance of the models. Firstly, we use the mean absolute error (MAE, see Eq.~\eqref{eq:mae}) to evaluate the deviation of the predicted potential energy of each particle from its actual value. The curly bracket ($\left<\cdots\right>$) refers to averaging over different samples. This is the most important performance index as it is directly associated with how well the model predicts the energy at the particle level.
    
    \begin{equation}
        \label{eq:mae}
        \left<|\delta \epsilon|\right> = \left<\sum_{i=1}^{n_\textrm{particle}}|\epsilon_\textrm{pred}^i-\epsilon_\textrm{true}^i|/n_\textrm{particle}\right>
    \end{equation}
    
    Secondly, the mean max error (MME, see Eq.~\eqref{eq:mme}) is used. It is used to evaluate the stability of the model. Even if the MAE is good, there is still a chance that a singular particle or a fraction of particles that deviate much more than MAE exists. The existence of such particles could result in misleading analysis at the system level since the mean value does not properly represent the energy distribution and degrade the reliability of the model in the end. MME can detect such anomalies at minimal cost by taking the maximum deviation in the system. We use four metrics to quantify and compare the performance of the models. Firstly, we use the mean absolute error (MAE, see Eq.~\eqref{eq:mae}) to evaluate the deviation of the predicted potential energy of each particle from its actual value. The curly bracket ($\left<\cdots\right>$) refers to averaging over different samples. This is the most important performance index as it is directly associated with how well the model predicts the energy at the particle level.
    
    \begin{equation}
        \label{eq:mme}
        \left<\max|\delta \epsilon|\right> = \left<\max_{i=1}^{n_\textrm{particle}}|\epsilon_\textrm{pred}^i-\epsilon_\textrm{true}^i|\right>
    \end{equation}
    
    Thirdly, the normalized MAE (NMAE, see Eq.~\eqref{eq:nmae}) is used for the consistent appreciation of the performance of systems at different energy levels. There are two problems comparing different systems: the energy scales and the distributions. Although the absolute index of how well the model predicts the energy can be represented by the MAE, comparing the performances of the model on different systems is difficult. A small deviation of energy in the system, which is on a lower energy scale, can be large enough for the system, which is on a higher energy scale. Another problem is that the distribution of energy can be different in each system. A sharp distribution makes a wrong prediction corresponding to the representative value (e.g., mean or most frequent value) bearable, while a relatively broad distribution makes such choice a big penalty in MAE. The NMAE reasonably deals with the above problem by dividing MAE with $\sum_{i=1}^{n_\textrm{particle}}|\epsilon_\textrm{true}^i-\bar{\epsilon}_\textrm{true}|$ where $\bar{\epsilon}_\textrm{true}$ is the average of true energy for each particle. Because the deviation of energy from its mean also grows with scale, the normalization can resolve the first problem. Furthermore, a broader energy distribution can get compensated in the denominator in NMAE.
    
    \begin{equation}
        \label{eq:nmae}
        \left<\mathcal{E}\right> = \left<\dfrac{\sum_{i=1}^{n_\textrm{particle}}|\epsilon_\textrm{pred}^i-\epsilon_\textrm{true}^i|}{\sum_{i=1}^{n_\textrm{particle}}|\epsilon_\textrm{true}^i-\epsilon_\textrm{true}|}\right>
    \end{equation}
    
    Lastly, the mean absolute percentage error (MAPE, see \eqref{eq:mape}) represents the error relative to the energy level of the system. This metric was prepared because the degree of error can be evaluated differently depending on the energy level of each system. Note that the definition of MAPE is slightly different from general MAPE in other literature. Usually, the denominator should be the actual energy of each particle, but since the energy of the particle can be zero, it can cause unexpected significant errors or unexpected values. Therefore, here we use a modified metric where the system-averaged energy replaces the denominator.
    
    \begin{equation}
        \label{eq:mape}
        \left<\mathcal{E}_\textrm{percent}\right> = \left<\dfrac{\sum_{i=1}^{n_\textrm{particle}}|\epsilon_\textrm{pred}^i-\epsilon_\textrm{true}^i|}{\sum_{i=1}^{n_\textrm{particle}}|\epsilon_\textrm{true}^i|}\right>
    \end{equation}
    The full list of data used in the following discussion can be checked in Table.~\ref{tab:results}
    
    \section{\label{sec:toy_model} Model Systems}

    \begin{figure*}[!htb]
    \includegraphics[width=7.0in, height=4.7in]{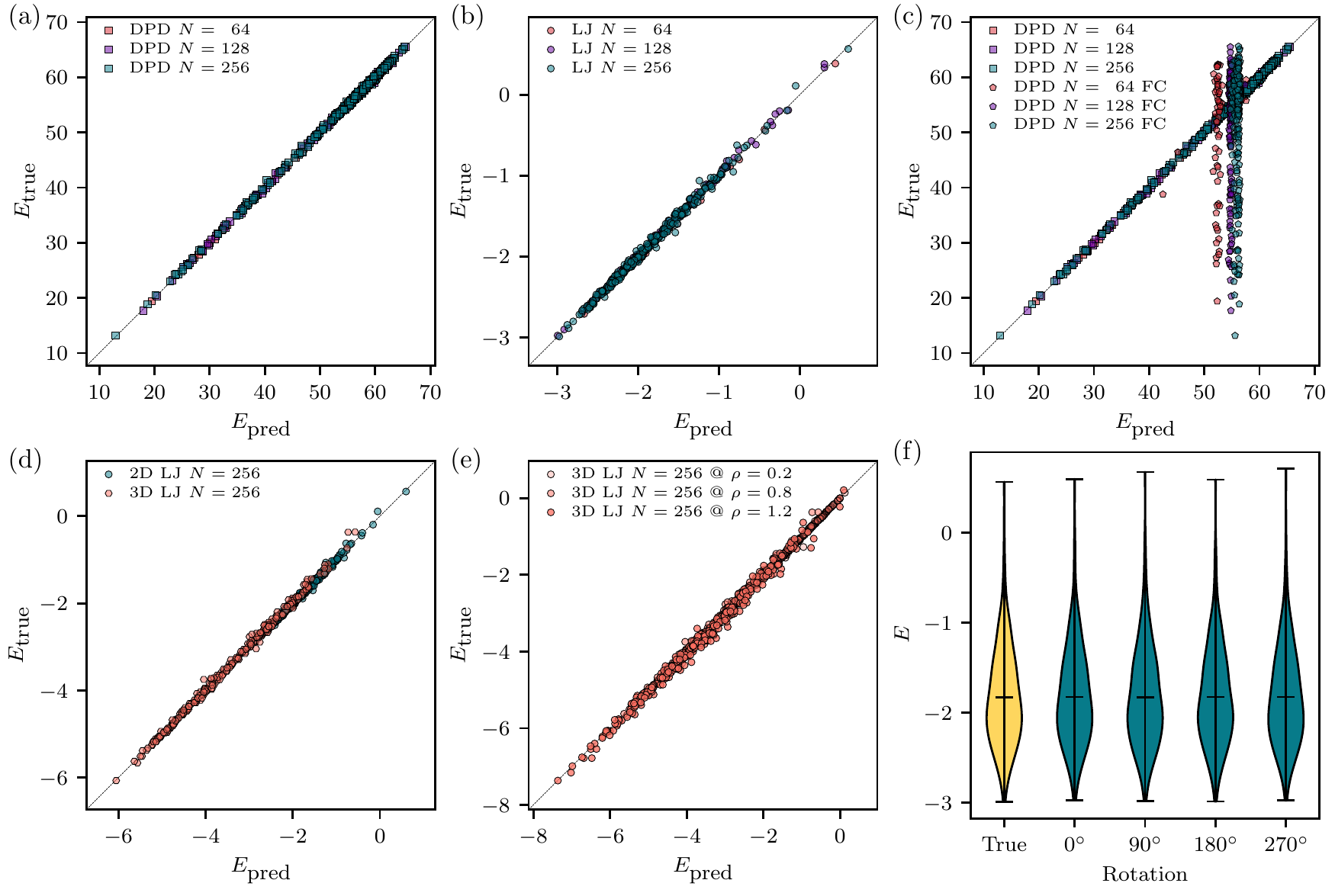}
    \caption{\label{fig:simple_system_result} The result of applying the model to (a,c) DPD and (b,d,e,f) LJ systems. (c) The comparison with the fully connected (FC) network shows the inability of FC to perform the task. The increase of the task's difficulty by increasing dimension (d) or density (e) does not significantly degrade the performance of the model. (f) The subsequent rotational transformation by $90^{\circ}$ makes a minimal distortion in the potential energy distribution.}
    \end{figure*}
    
    We start the discussion by demonstrating the performance of our model in a simple system. A large fraction of potential energy functions is given in the form of a pair potential function, where the key to estimating the pair potential energy is the interparticle distance. Among popular simple potentials, the conservative energy of Dissipative Particle Dynamics (DPD) \cite{groot1997dissipative} has similar forms to the interparticle distance itself as shown in Eq.~\eqref{eq:dpd}. As the corresponding force is Hookean force proportional to the interparticle distance, DPD conservative energy is one of the simplest potential functions.
    
        \begin{equation}
        \label{eq:dpd}
        U_{ij}\left(r\right)=
        \left\{\!\begin{aligned}
        A\bigl(r_{ij}&-r_c\bigr)^2 && \textrm{if} \ r_{ij}<r_c \\
        &0 && \textrm{if} \ r_{ij}\geq r_c
        \end{aligned}\right.
    \end{equation}
    
    We used energy coefficient $A{=}50$ and cutoff $r_c{=}2.5$. The number density is $\rho=0.8$, larger than typical DPD simulations to ensure sufficient neighbors for each particle. Three sets of fixed boundary DPD systems ($N{=}64,128,256$) in $2$D are prepared. We run PBC MD simulation in canonical ensemble (NVT) at a unit temperature and recalculate the energy under fixed boundary in the postprocessing. Note that all MD simulations, which only include pair potentials, were performed using HOOMD, and dimensionless Lennard-Jones (LJ) unit is used unless otherwise specified. The results are shown in FIG.~\ref{fig:simple_system_result}(a). The MAE is $\sim {O}\left(10^{-1}\right)$ and it is $<1\%$ MAPE. The MAE increases as the system size increases, but it is attributed to the increased energy level due to the fixed boundary. The MME is also $\sim {O}\left(10^{-1}\right)$ which indicates the model is stable enough.
    
    As the model worked very well in the DPD system, a more complex potential was tested, which is the LJ particle system. LJ potential, which is one of the most widely used potentials, has a rapidly diverging repulsive core at $r=0$ (see Eq.~\eqref{eq:lj}). We set $\epsilon$ and $\sigma$ to unity, respectively. As in the DPD case, we prepared three system sizes ($N{=}64,128,256$) in $2$D at $\rho=0.8$ under fixed boundary condition. The energy is calculated with the same procedure as in the DPD system. Note that, for the simplicity of discussion, the same parameters, density, and preparation protocols are used for LJ systems unless otherwise specified. The result is shown in FIG.~\ref{fig:simple_system_result}(b).
        
    \begin{equation}
    \label{eq:lj}
        U_{ij}\left(r\right)=
        \left\{\!
        \begin{aligned}
        \epsilon\bigg[\Bigl(\dfrac{\sigma}{r_{ij}}\Bigr)^{12}&-\Bigl(\dfrac{\sigma}{r_{ij}}\Bigr)^{6}\bigg] - U\left(r_c\right) && \textrm{if} \  r_{ij}<r_c \\
        &0 && \textrm{if} \ r_{ij}\geq r_c
        \end{aligned}
        \right.
    \end{equation}
    
    Unlike DPD, learning distance alone is not enough to predict LJ potential. Suitable expressive power with which learned distance can be internally converted to LJ potential is needed. Our model shows MAE $~O(10^{-2})$ and corresponding MAPE is $\sim 1\%$. The MME shows $~O(10^{-1})$ which is slightly larger than DPD but still acceptable because the NMAEs of both LJ and DPD are $~O(10^{-2})$. It indicates that they have a similar level of predictability when scale and distribution are taken into account. The moderate increase of MAE with the increasing number of particles also appeared for the same reason as in DPD systems. 
    
    Before further analysis, we provide how the conventional fully connected network (FC) works with DPD systems. As shown in FIG.~\ref{fig:simple_system_result}(c), FC completely fails to predict the energy even with the most simple potential tested. One interesting observation is that FC remains at its best strategy, predicting all particles with uniform energy close to the system's mean energy. When all particles have the mean energy of the system, the denominator in the NMAE formula is equal to MAE. Thus, NMAE can be approximately interpreted as how well the model predicts the energy compared to FC.
    
    We further tested the model by changing the dimension of the LJ system from $2$D to $3$D, and the result is shown in FIG.~\ref{fig:simple_system_result}(d). Preparation protocols or parameters are the same except that we set the density $\rho=0.8/\sigma^3$ in $3$D while in 2D we used $\rho=0.8/\sigma^2$. Although the MAE of $3$D is a bit larger than $2$D due to the increased dimensionality of the task, the model still keeps MAE in the same order.
    
    The key to performing the task is to estimate contributions from the neighboring particles. It means that the number of neighbors affects the performance by increasing the dimensionality of the task. As a sanity check, we built three fixed boundary $3$D LJ systems at different densities ($\rho=0.2,0.8,1.2$). As shown in FIG.~\ref{fig:simple_system_result}(e), the more dense system has a larger MAE. However, even at $\rho=1.2$, the MAE is $\sim O(10^{-1})$, and it corresponds to the MAPE $\sim4\%$, which is reasonably acceptable for most of the purposes.
    
    Finally, we investigated how much the model is robust against rotations. In an amorphous phase, the system should obey rototranslational invariance. Specifically, in a cubic simulation box, the system energy should be four-fold invariant under the rotation of $90^{\circ}$, $180^{\circ}$, and $270^{\circ}$; if there is a systematic bias of energy under any of these rotations, the model fails to capture the symmetry of the system. However, as shown in FIG.~\ref{fig:simple_system_result}(f), the prediction for each rotation is almost the same as the one for the original configuration. The result shows that even though the model is not designed to obey the rotational symmetry, the model works just fine for the rotated configurations. This might be due to our model's ability to understand the underlying physics and provides evidence that we avoided overfitting to a handful of snapshots.
    
    We have demonstrated that, for simple but physically meaningful systems, our model works surprisingly well so far. In the next section, our model is applied to more complex and general systems to prove its generalizability.

\section{\label{sec:extended} Extended systems}
    \subsection{Periodic Boundary Condition}
    
    \begin{figure*}[!htb]
    \includegraphics[width=7.0in, height=4.7in]{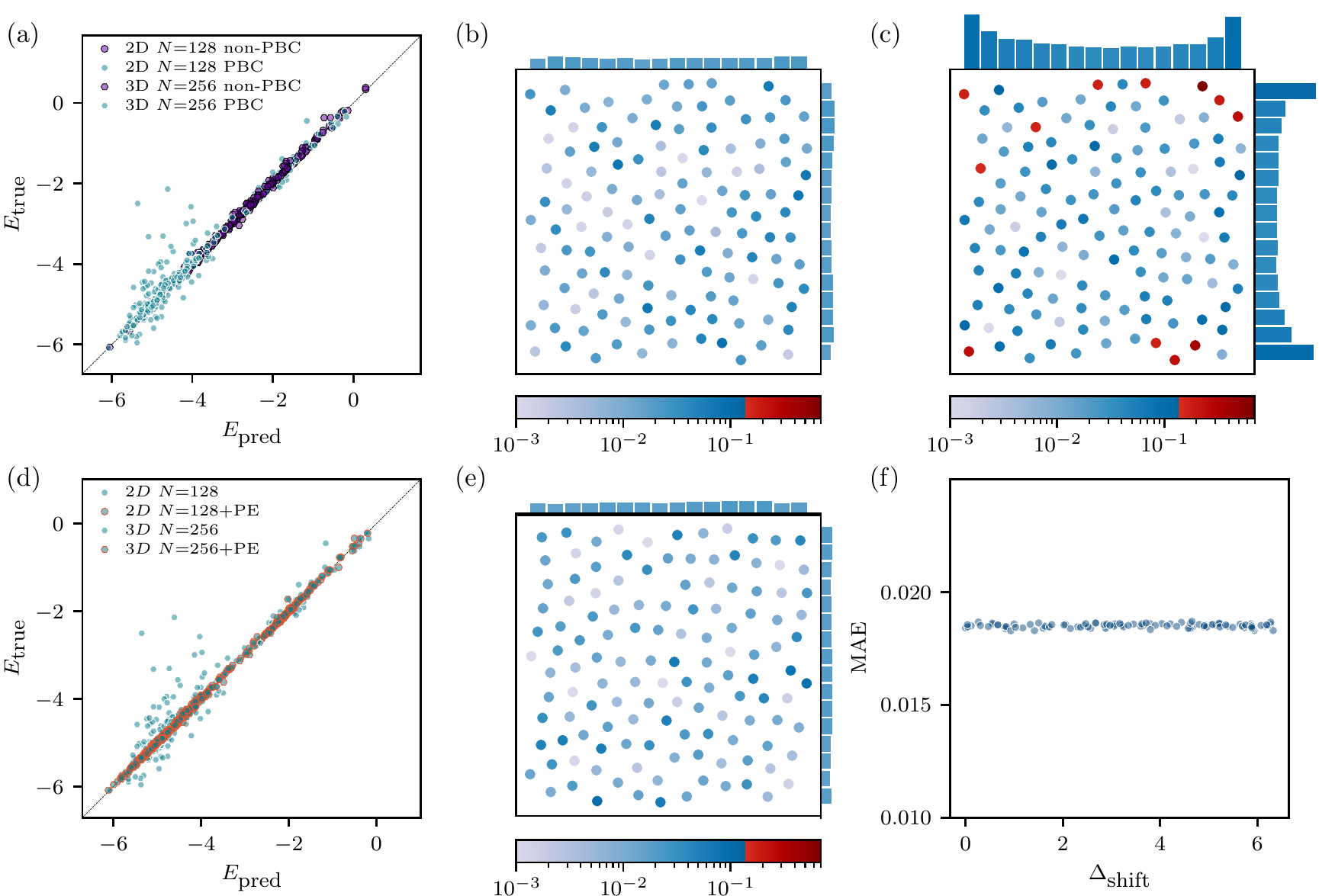}
    \caption{\label{fig:pbc_system_result} The comparisons of (a) fixed boundary and PBC systems and (d) PBC and PBC+PE systems show the effect of PE. The source of the significant deviation in the predicted energy from the actual one can be explained by the relatively poor performance of the model to deal with the particles near the boundary. The unsigned energy deviation for (b) fixed boundary and (c) PBC systems clearly show the difference. (d) The large deviations observed in PBC systems are removed in the PBC+PE system, and (e) the spatially inhomogeneous prediction is also alleviated. Note that bar plots located above and at the right of each (b),(c), and (d) are spatially resolved MAE, and the particle color (energy deviation) follows the color bar below each plot. (f) Any shifts with random distance ($\Delta_\textrm{shift}$) in an arbitrary direction give almost no fluctuation in the MAE}
    \end{figure*}

    The first complexity that we consider is PBC. In the earlier section, we only tested fixed boundary systems. However, many molecular sciences feature PBC to fill the astronomical gap between the simulation system size and continuum bulk size. Therefore, the model should properly handle PBC for practical use.
    
    FIG.~\ref{fig:pbc_system_result}(a) shows the result of fixed boundary and PBC systems in both $2$D and $3$D using the same model architecture as in FIG.~\ref{fig:simple_system_result}. The evident deviation for the PBC system compared to the fixed boundary system is clearly observed. Although the MAE is higher than the same systems other than the PBC condition, the MAE itself is not the worst. Moreover, It can be seen that the MME is ten times higher than MAE in both $2$D and $3$D. It means that a fraction of particles disturbs the stability of prediction by deviating more than the rest of the particles. The origin of the deviation can be found by comparing FIG.~\ref{fig:pbc_system_result}(b) and (c). In $2$D fixed boundary system (FIG.~\ref{fig:pbc_system_result}(b)), the deviation of each particle is spatially homogeneous. On the contrary, in the $2$D PBC system (FIG.~\ref{fig:pbc_system_result}(c)), there is a clear trend that the particles with large deviations are at the boundary. Thus, we can conclude that the pristine model is weak at predicting the particle near the boundary under PBC.
    
    How come the model has such weakness? Most LJ pair interactions do not cross the boundary, and the interactions are between spatially close particles. Thus, the model learns such nature very well, as we see in the previous section. However, a small fraction of the interactions is across the boundary, which acts as a noise. Thus, the learning for those interactions progresses poorly. Furthermore, as the model already learned that distant particles do not interact, it is hard for the model to distinguish pairs farther away from each other particles.
    
\begin{figure}[!htb]
    \includegraphics[width=3.4in, height=3.4in]{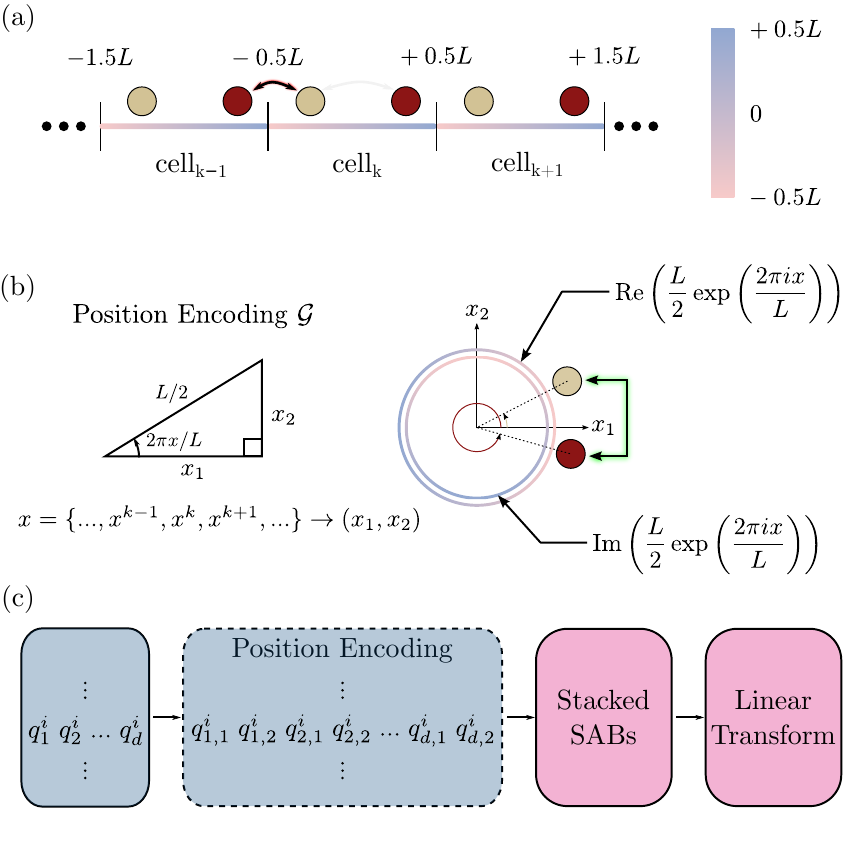}
    \caption{\label{fig:position_encoding} The schematic diagram of position encoding (PE) for PBC is shown. Two particles interact across the boundary for the minimum image convention, which requires a high association between numerically faraway particles at the blue and pink coordinates, as shown in (a). (b) By using PE, $x_1$ and $x_2$ are continuous in each coordinate. The PE recovers the rule that adjacent coordinates interact with each other as in fixed boundary systems without complexity, such as the minimum image convention. (c) The final network architecture with the PE implemented is shown.}
    \end{figure}

    As shown in the FIG.~\ref{fig:position_encoding}(a), the fundamental problem is that counting numerically nearby particles in the coordinate space does not ensure all pairs of interactions because of the existence of numerical cliffs at the periodic boundaries. We introduce a new method called \textit{position encoding} (PE, see FIG.~\ref{fig:position_encoding}(b)). We split each dimension into two using the following transformation:
    
    \begin{equation}
    \label{eq:pe}
    \begin{aligned}
        \mathcal{G} &: x \longrightarrow \left(x_1,x_2\right)\\
        x_1 &= L/2 \cos \left(\dfrac{2 \pi i x}{L}\right)\\
        x_2 &= L/2 \sin \left(\dfrac{2 \pi i x}{L}\right)
    \end{aligned}    
    \end{equation}
    
    \noindent where $L$ is the periodic box length. The transformation is reversible, so any coordinate is uniquely mapped to a pair of coordinates in the transformed coordinates. We leveraged the periodicity of trigonometric functions to encode the periodic character. This reminds us of the complex number $z=L/2 \exp(2 \pi i x / L)$$=L/2\cos(2 \pi i x / L) + i\sin(2 \pi i x / L) $$=x_1+ix_2$. With the PE, the $(x_1,x_2){-}\textrm{coordinate space}$ is continuous in number as shown in the right schematic of FIG.~\ref{fig:position_encoding}(b). Note that the position encoding in this work is distinct from the \textit{positional encoding} in Transformer \cite{vaswani2017attention} which provided the relative position of word token in a sentence.
    
    With the newly introduced PE, we can recover the performance that is lost in PBC systems. As shown in FIG.~\ref{fig:pbc_system_result}(d), the comparison of the models with and without PE clearly shows the superiority of PE over the pristine model. Moreover, spatially inhomogeneous prediction also is hard to observe after the introduction of PE (see FIG.~\ref{fig:pbc_system_result}(e)). The MAE of the PBC system with PE is the same as or even lower than the fixed boundary system. The fact that the system could have lower MAE might be puzzling, but, considering that the PBC system is more spatially homogeneous due to the presence of boundary particles near the fixed boundary, the result has no contradiction. The NMAE is also consistent with it since the fixed boundary system has a broader energy distribution due to boundary particles. Thus, the NMAEs of fixed boundary systems are lower than the PBC+PE counterparts even though the MAEs of fixed boundary systems are the same or higher. Note that we will use PE for all PBC systems in the rest of this paper.
    
    As a follow-up test for the stability of the model against rotation, we tried to measure the stability of the model against translation. We shifted randomly sampled $256$ configurations of $2$D LJ PBC+PE system ($N=128$) with the randomly chosen shift distances ($\Delta_\textrm{shift}$) and directions. The result is astounding as shown in FIG.~\ref{fig:pbc_system_result}(f). During sequential translation, the MAE of the model hardly deviated. 
    
    Based on the results of rotation and translation tests, we can argue that, even though the model does not have an inherent structure that strictly follows rotational and translational invariance, our model complies with the constraints to a satisfactory level.
    
    \subsection{Higher-order Interaction}
    
    While the most common form of interaction in classical MD is the pair potential, the origin of the interaction between particles is a complex interplay of wave functions whose nature is $n$-body. Nevertheless, the computational cost gap between pair potential and interaction by quantum mechanical calculation is so large that only small-size systems can be simulated in a reasonable time using the latter. Thus, researches to embrace quantum mechanical nature within classical MD with the minimum expense of computational cost still actively continue. The most straightforward idea is designing a potential which directly includes $n$-body ($n>2$) interaction, such as $n=3$ Stillinger-Weber \cite{stillinger1985computer} and Tersoff \cite{tersoff1988new} potentials. Inspired by Set Transformer \cite{lee2019set} in which higher-order interaction can be encoded, we tested the ability of our model to capture the underlying higher-order interaction ($n=3$).
    
        \begin{equation}
    \label{eq:sw}
    \begin{split}
        U&=\sum_{i}\sum_{j>i}U_{ij} + \sum_{i}\sum_{j\neq i}\sum_{k}U_{ijk} \\
        U_{ij}( r_{ij} )&=A_{ij}\epsilon_{ij}\bigg[B_{ij}\Bigl(\dfrac{\sigma}{r_{ij}}\Bigr)^{p_{ij}}-\Bigl(\dfrac{\sigma}{r_{ij}}\Bigr)^{q_{ij}}\bigg] \\ 
        &\phantom{=} \times \exp\bigg(\dfrac{\sigma_{ij}}{r_{ij}-a_{ij}\sigma_{ij}}\bigg) \\
        U_{ijk}\big(r_{ij}, r_{ik}, \theta_{ijk}\big) &= \lambda_{ijk}\epsilon_{ijk}\big[\cos{\theta_{ijk}} - \theta_{0, ijk}\big]^2 \\
        &\phantom{=} \times \exp\bigg(\dfrac{\gamma_{ij}a_{ij}}{r_{ij}-a_{ij}\sigma_{ij}}\bigg)\\ 
        &\phantom{=} \times \exp\bigg(\dfrac{\gamma_{ik}a_{ik}}{r_{ik}-a_{ik}\sigma_{ik}}\bigg)
    \end{split}
    \end{equation}
    
    Among the above mentioned $3$-body potentials, we tested Stillinger-Weber potential (see Eq.~\eqref{eq:sw}) for the test. We prepared three systems ($N=216$) with different tetrahedrality parameters, $\lambda$ controlling the relative strength of the $3$-body energy over pairwise energy. The systems were developed to simulate germanium ($\lambda=20.0$) \cite{bhat2007vitrification}, silicon ($\lambda=21.0$) \cite{stillinger1985computer}, and monatomic water ($\lambda=23.15$) \cite{molinero2009water}. For other parameters, we used $\epsilon=\SI{6.189}{\kcal\per\mole}$, $\sigma=\SI{2.3925}{\angstrom}$, $\gamma=1.20$, $\cos\theta_0=-1/3$, and $A=\SI{7.049556277}{}$, $B=\SI{0.6022245584}{}$, $p=4$, $q=0$, and $a=1.8$ throughout the test. The MD simulations were performed in NVT at $\rho=\SI{1.0}{\gram\per\cubic\centi\metre}$ and $T=\SI{300}{\kelvin}$ using LAMMPS.
    
    \begin{figure}[!htb]
    \includegraphics[width=2.38in, height=2.38in]{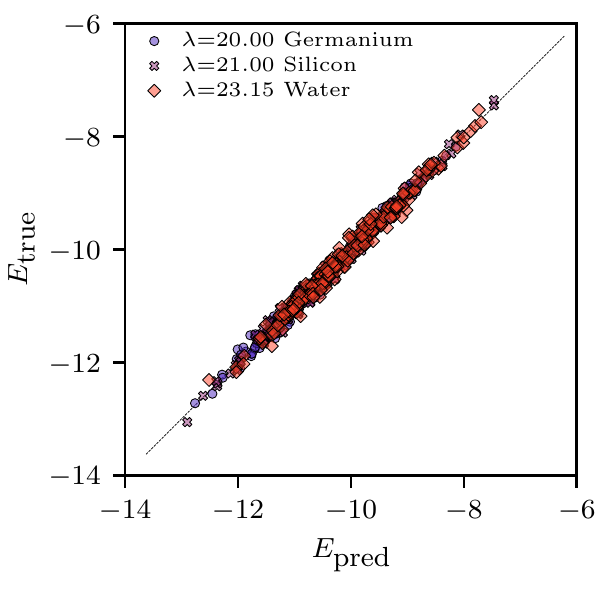}
    \caption{\label{fig:stillinger_weber} The prediction-true energy plots for Stillinger-Weber potential with different tetrahedrality parameters ($\lambda$)}
    \end{figure}
    
    As shown in FIG.~\ref{fig:stillinger_weber}, the model works well for all three tetrahedrality parameters. The MAEs are ${\sim}O(10^{-2}) \SI{}{\kcal\per\mole}$ and corresponding MMEs and NMAEs are ${\sim}O(10^{-1}) \SI{}{\kcal\per\mole}$. The MAE corresponds to the MAPE $<1\%$. Similar MAPE for the Stillinger-Weber system to that of the pair potential system indicates that the model possibly works for both pair potential and $3$-body potential at a similar level of precision.
    
    \subsection{Binary}
    
    The last complexity that we discuss is the handling of heterogeneous particles. Simulations involving multiple species of particles are prevalent. Giving the model the ability to grasp particle species greatly improve its use. The most general approach is to construct separate local networks for each case \cite{behler2007generalized,smith2017ani}. Specifically, for the binary system of A- and B-type particles, we can construct two self-attention networks with the additional attention network using MAB(A, B) (see Eq.~\eqref{eq:MAB}). However, the idea has three major disadvantages. Firstly, as the number of species ($n$) increases, the required number of networks quadratically increases. Specifically, the networks for $n$ homo-species interaction and the ones for $_nC_r$ hetero-species interaction should be prepared. Secondly, when composition changes, the model should be reformulated accordingly. For instance, if the model is trained for a system ($N=100$) in which $N_A=50$ and $N_B=50$, The same network cannot be used for composition $N_A=51$ and $N_B=49$. Lastly, deriving the prediction out of outputs from each local network is nontrivial. The local network outputs should be properly processed for the prediction. The way to combine and transform into the final form is neither straightforward nor unique.
    
    We came up with an exciting method to include multicomponent systems with minimal modification. The method uses indistinguishable particles in place of distinguished particles by introducing an extra dimension. When types are assigned to particles, A parameter of a type is expected to be different from that of another, and that makes each type distinguishable. Instead of such "coloring," we can think of identical particles with different coordinates in the \textit{extra dimension}.
    
\begin{figure}[!htb]
        \includegraphics[width=2.38in, height=4.76in]{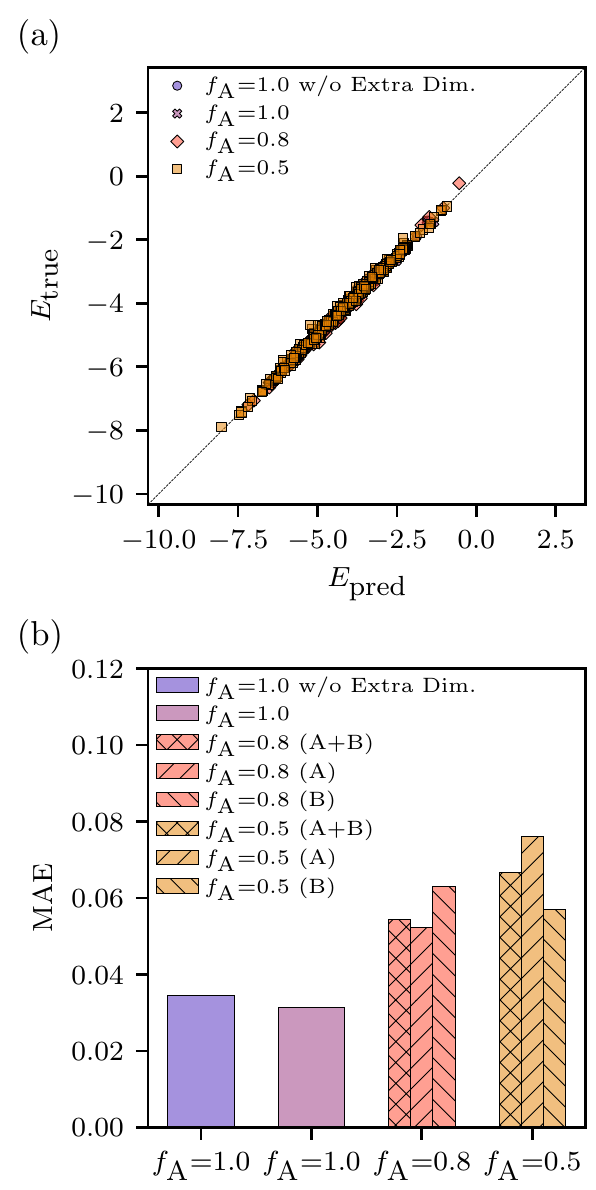}
        \caption{\label{fig:binary_result} (a) The prediction-true energy plot for binary LJ systems with different A-type particle fractions ($f_\textrm{A}$). (b) Comparison of the systems given in (a). The extra dimension does not increase the error, and the breakdown of performance for particle types illustrates that the performance of the model does not depend on the type of potentials}
    \end{figure}

    To illustrate the effectiveness of the method, we prepared three 3D LJ systems ($N=256$). In each system, we vary the proportions of A-type particles ($f_\textrm{A}=1.0$, $f_\textrm{A}=0.8$, and $f_\textrm{A}=0.5$, respectively). The set of parameters are taken from the Kob-Andersen mixture \cite{kob1995testing}. In detail, $\sigma_{AA}=1.0$, $\sigma_{BB}=0.88$, $\sigma_{AB}=0.8$, $\epsilon_{AA}=1.0$, $\epsilon_{BB}=0.5$, and $\epsilon_{AB}=1.5$ are used. In the extra dimension, the A- and B-type particles are located at $0$ and $1$, respectively. As shown in FIG.~\ref{fig:binary_result}(a), the predicted energies follow the actual energies very precisely for all cases with MAPE $\sim 1\%$. Furthermore, the performance of the model does not depend on the types of particles as shown in the type-breakdown analysis in FIG.~\ref{fig:binary_result}(b).
    
    Note that the MAE gets slightly higher as the fraction of B-type particles increases. This is not because we have the extra dimension. When two $f_\textrm{A}=1.0$ systems with and without extra dimension are compared in FIG.~\ref{fig:binary_result}(b), just inserting the extra dimension does not increase the MAE. The source of the increase is the increase in the dimensionality of the task. The inclusion of $N_\textrm{A}$ A-type particles among $N$ particles increases the dimensionality of task by $N_\textrm{A}!(N-N_\textrm{A})!$ compared to identical $N$ particles.

\begin{table*}[!htb]

\caption{Mean absolue error ($\left<|\delta E|\right>$), mean max error ($\left<\textrm{max}|\delta E |\right>$), normalized mean absolute error ($\left<|\mathcal{E}|\right>$), and mean absolute percent error ($\left<|\mathcal{E}_\textrm{percent}|\right>$) for all system tested. Note that the number of samples for the data ($N_\textrm{sample}$) is $128$. The units of $\left<|\delta E|\right>$ and $\left<\textrm{max}|\delta E |\right>$ are given in Lennard-Jones dimensionless energy unit except for Stillinger-Weber (SW) which is given in \SI{}{\kcal\per\mole}. The units of $\left<|\mathcal{E}|\right>$ and $\left<|\mathcal{E}_\textrm{percent}|\right>$ are dimensionless fraction and $\%$, respectively.}
\begin{tabularx}{\linewidth}{ZZZQGGGG}\toprule\toprule
\multicolumn{4}{c}{Condition} & \multicolumn{1}{c}{$\left<|\Delta E|\right>$} & \multicolumn{1}{c}{$\left<\textrm{max}|\delta E|\right>$} & \multicolumn{1}{c}{$\left<|\mathcal{E}|\right>$} & \multicolumn{1}{c}{$\left<|\mathcal{E}_\textrm{percent}|\right>$}\\
\midrule
{$2$D} & {DPD} & {No PBC} & $N=64$ & $0.111 {\pm} 0.014$ & $0.366 {\pm} 0.094$ & $0.011 {\pm} 0.001$ & $0.233 {\pm} 0.030$  \\
& & & $N=128$ &$0.186 {\pm} 0.014$ & $0.672 {\pm} 0.120$ & $0.021 {\pm} 0.002$ & $0.370 {\pm} 0.028$  \\
& & & $N=256$ &$0.207 {\pm} 0.012$ & $0.867 {\pm} 0.204$ & $0.028 {\pm} 0.002$ & $0.394 {\pm} 0.022$  \\\cmidrule{2-2}
& DPD (FC) & & $N=64$ &$9.298 {\pm} 0.631$ & $35.810 {\pm} 3.840$ & $0.897 {\pm} 0.041$ & $19.589 {\pm} 1.410$  \\
& & & $N=128$ & $7.951 {\pm} 0.364$ & $38.579 {\pm} 3.580$ & $0.889 {\pm} 0.022$ & $15.781 {\pm} 0.757$  \\
& & & $N=256$ & $6.504 {\pm} 0.257$ & $39.183 {\pm} 3.531$ & $0.885 {\pm} 0.014$ & $12.404 {\pm} 0.503$  \\\cmidrule{2-2}
& LJ & & $N=64$ & $0.022 {\pm} 0.004$ & $0.124 {\pm} 0.067$ & $0.043 {\pm} 0.007$ & $1.320 {\pm} 0.279$  \\
& & & $N=128$ & $0.020 {\pm} 0.003$ & $0.121 {\pm} 0.041$ & $0.040 {\pm} 0.005$ & $1.149 {\pm} 0.187$  \\
& & & $N=256$ & $0.030 {\pm} 0.004$ & $0.249 {\pm} 0.119$ & $0.062 {\pm} 0.008$ & $1.644 {\pm} 0.277$  \\\cmidrule{3-3}
& & PBC & $N=128$ & $0.052 {\pm} 0.009$ & $0.466 {\pm} 0.273$ & $0.131 {\pm} 0.022$ & $2.642 {\pm} 0.520$  \\\cmidrule{3-3}
& & PBC (PE) & $N=128$ & $0.019 {\pm} 0.002$ & $0.127 {\pm} 0.049$ & $0.046 {\pm} 0.005$ & $0.936 {\pm} 0.130$  \\\cmidrule(l){1-1} \cmidrule{3-3} \cmidrule{2-2}
{$3$D} & {LJ} & {No PBC} & {$N=256$\newline$\left(\rho=0.2\right)$} & $0.028 {\pm} 0.004$ & $0.275 {\pm} 0.133$ & $0.044 {\pm} 0.006$ & $2.160 {\pm} 0.252$  \\
& & &  $N=256$\newline$\left(\rho=0.8\right)$ & $0.044 {\pm} 0.005$ & $0.324 {\pm} 0.098$ & $0.044 {\pm} 0.005$ & $1.269 {\pm} 0.159$  \\
& & &  {$N=256$\newline$\left(\rho=1.2\right)$} & $0.139 {\pm} 0.012$ & $0.670 {\pm} 0.156$ & $0.118 {\pm} 0.011$ & $3.578 {\pm} 0.289$  \\\cmidrule{3-3}
& & PBC & $N=256$ &$0.164 {\pm} 0.018$ & $1.779 {\pm} 0.615$ & $0.306 {\pm} 0.032$ & $3.494 {\pm} 0.396$  \\\cmidrule{3-3}
& & PBC (PE) & $N=256$ & $0.034 {\pm} 0.003$ & $0.223 {\pm} 0.071$ & $0.063 {\pm} 0.006$ & $0.714 {\pm} 0.074$  \\\cmidrule{2-2}
& {SW\newline$\left(N=216\right)$} & & {Germanium\newline$\left(\lambda=20.00\right)$} & $0.081 {\pm} 0.005$ & $0.323 {\pm} 0.049$ & $0.112 {\pm} 0.008$ & $0.780 {\pm} 0.049$  \\
& & & {Silicon\newline$\left(\lambda=21.00\right)$} & $0.084 {\pm} 0.005$ & $0.335 {\pm} 0.054$ & $0.116 {\pm} 0.008$ & $0.818 {\pm} 0.048$  \\
& & & {Water\newline$\left(\lambda=23.15\right)$} & $0.092 {\pm} 0.005$ & $0.371 {\pm} 0.069$ & $0.127 {\pm} 0.010$ & $0.917 {\pm} 0.055$  \\\cmidrule{2-2}
& {LJ\newline(Extra Dim.)}& &  {$N=256$\newline$\left(f_A=1.0\right)$} & $0.032 {\pm} 0.003$ & $0.209 {\pm} 0.067$ & $0.059 {\pm} 0.006$ & $0.672 {\pm} 0.067$  \\
& & & {$N=256$\newline$\left(f_A=0.8\right)$} & $0.055 {\pm} 0.004$ & $0.415 {\pm} 0.165$ & $0.080 {\pm} 0.006$ & $1.203 {\pm} 0.101$  \\
& & & {$N=256$\newline$\left(f_A=0.5\right)$} & $0.067 {\pm} 0.007$ & $0.483 {\pm} 0.158$ & $0.059 {\pm} 0.007$ & $1.582 {\pm} 0.177$  \\\bottomrule\bottomrule
\end{tabularx}
\label{tab:results}
\end{table*}

\section{\label{sec:future} Conclusion and Future work}

    We developed a multiobjective machine learning model that can predict each particle's potential energy given only the coordinates of the particle. We have demonstrated the performance of our model for a variety of systems with in-depth analysis and discussion. We introduced a new method (PE) to deal with the PBC. Furthermore, we demonstrated that the model could learn higher-order interaction through the examples of Stillinger-Weber potential. We also showed the ability of the model to deal with the binary system with the simple idea of extra dimension. One more interesting result is that the model works very well for rotational and translational transformations without us taking special treatment for it. All-in-all, we developed a model very versatile for many of the molecular simulations. The presented network and ideas can also be implemented in any particle-based simulation.
    
    In future work, we aim to develop a model that can deal with covalent interactions such as bond or angle potentials, which are out of scope for this work. The long-range interaction such as Ewald summation or particle-mesh methods is also one of the quests remaining for future work. As we only tested $3$-body potential as the proof-of-concept for higher-order interaction, we will further extend this concept to higher-order interactions. We are focusing on lowering the computation cost of DFT-level interaction to that of typical machine learning level. 
    
    Although the task we showed in this work predicts the potential energy at the particle level, the method can be generally applied to other properties. Indeed, the particle-level prediction has been widely used recently. For example, \textit{soft} particles \cite{cubuk2015identifying, schoenholz2017relationship}, local structure prediction \cite{boattini2019unsupervised}, liquid- or gas-like particles \cite{ha2019universality} are all based on machine learning. Note that our model relies only on the repetition of SAB blocks without resorting to complex model architecture or exhaustive feature engineering. Further tuning of the architecture or input vector might significantly enhance the already satisfactory prediction performances, to open a new avenue for highly accurate approximation of the potential energy surface with reasonable computational cost., so the method can be implemented and applied to the above existing machine learning works. The method has superiority over the existing methods, which take neighboring particles as an additional input in that attention mechanism proves the correlation between all particles. It means the model can be expected to work very well with deeply correlated and cooperative systems like glass.
    
\section{ACKNOWLEDGEMENTS}

This work was supported by the National Research Foundation of Korea (NRF) grant funded by the Korea government (MSIT) (No. NRF-2018M3D1A1058633, 2019R1A2C1085081, and 2021M3H4A6A01041234).


\begin{thebibliography}{46}%
\makeatletter
\providecommand \@ifxundefined [1]{%
 \@ifx{#1\undefined}
}%
\providecommand \@ifnum [1]{%
 \ifnum #1\expandafter \@firstoftwo
 \else \expandafter \@secondoftwo
 \fi
}%
\providecommand \@ifx [1]{%
 \ifx #1\expandafter \@firstoftwo
 \else \expandafter \@secondoftwo
 \fi
}%
\providecommand \natexlab [1]{#1}%
\providecommand \enquote  [1]{``#1''}%
\providecommand \bibnamefont  [1]{#1}%
\providecommand \bibfnamefont [1]{#1}%
\providecommand \citenamefont [1]{#1}%
\providecommand \href@noop [0]{\@secondoftwo}%
\providecommand \href [0]{\begingroup \@sanitize@url \@href}%
\providecommand \@href[1]{\@@startlink{#1}\@@href}%
\providecommand \@@href[1]{\endgroup#1\@@endlink}%
\providecommand \@sanitize@url [0]{\catcode `\\12\catcode `\$12\catcode
  `\&12\catcode `\#12\catcode `\^12\catcode `\_12\catcode `\%12\relax}%
\providecommand \@@startlink[1]{}%
\providecommand \@@endlink[0]{}%
\providecommand \url  [0]{\begingroup\@sanitize@url \@url }%
\providecommand \@url [1]{\endgroup\@href {#1}{\urlprefix }}%
\providecommand \urlprefix  [0]{URL }%
\providecommand \Eprint [0]{\href }%
\providecommand \doibase [0]{https://doi.org/}%
\providecommand \selectlanguage [0]{\@gobble}%
\providecommand \bibinfo  [0]{\@secondoftwo}%
\providecommand \bibfield  [0]{\@secondoftwo}%
\providecommand \translation [1]{[#1]}%
\providecommand \BibitemOpen [0]{}%
\providecommand \bibitemStop [0]{}%
\providecommand \bibitemNoStop [0]{.\EOS\space}%
\providecommand \EOS [0]{\spacefactor3000\relax}%
\providecommand \BibitemShut  [1]{\csname bibitem#1\endcsname}%
\let\auto@bib@innerbib\@empty
\bibitem [{\citenamefont {Jorgensen}\ and\ \citenamefont
  {Tirado-Rives}(1988)}]{jorgensen1988opls}%
  \BibitemOpen
  \bibfield  {author} {\bibinfo {author} {\bibfnamefont {W.~L.}\ \bibnamefont
  {Jorgensen}}\ and\ \bibinfo {author} {\bibfnamefont {J.}~\bibnamefont
  {Tirado-Rives}},\ }\bibfield  {title} {\bibinfo {title} {The opls [optimized
  potentials for liquid simulations] potential functions for proteins, energy
  minimizations for crystals of cyclic peptides and crambin},\ }\href@noop {}
  {\bibfield  {journal} {\bibinfo  {journal} {Journal of the American Chemical
  Society}\ }\textbf {\bibinfo {volume} {110}},\ \bibinfo {pages} {1657}
  (\bibinfo {year} {1988})}\BibitemShut {NoStop}%
\bibitem [{\citenamefont {Jorgensen}\ \emph {et~al.}(1996)\citenamefont
  {Jorgensen}, \citenamefont {Maxwell},\ and\ \citenamefont
  {Tirado-Rives}}]{jorgensen1996development}%
  \BibitemOpen
  \bibfield  {author} {\bibinfo {author} {\bibfnamefont {W.~L.}\ \bibnamefont
  {Jorgensen}}, \bibinfo {author} {\bibfnamefont {D.~S.}\ \bibnamefont
  {Maxwell}},\ and\ \bibinfo {author} {\bibfnamefont {J.}~\bibnamefont
  {Tirado-Rives}},\ }\bibfield  {title} {\bibinfo {title} {Development and
  testing of the opls all-atom force field on conformational energetics and
  properties of organic liquids},\ }\href@noop {} {\bibfield  {journal}
  {\bibinfo  {journal} {Journal of the American Chemical Society}\ }\textbf
  {\bibinfo {volume} {118}},\ \bibinfo {pages} {11225} (\bibinfo {year}
  {1996})}\BibitemShut {NoStop}%
\bibitem [{\citenamefont {Wang}\ \emph {et~al.}(2004)\citenamefont {Wang},
  \citenamefont {Wolf}, \citenamefont {Caldwell}, \citenamefont {Kollman},\
  and\ \citenamefont {Case}}]{wang2004development}%
  \BibitemOpen
  \bibfield  {author} {\bibinfo {author} {\bibfnamefont {J.}~\bibnamefont
  {Wang}}, \bibinfo {author} {\bibfnamefont {R.~M.}\ \bibnamefont {Wolf}},
  \bibinfo {author} {\bibfnamefont {J.~W.}\ \bibnamefont {Caldwell}}, \bibinfo
  {author} {\bibfnamefont {P.~A.}\ \bibnamefont {Kollman}},\ and\ \bibinfo
  {author} {\bibfnamefont {D.~A.}\ \bibnamefont {Case}},\ }\bibfield  {title}
  {\bibinfo {title} {Development and testing of a general amber force field},\
  }\href@noop {} {\bibfield  {journal} {\bibinfo  {journal} {Journal of
  computational chemistry}\ }\textbf {\bibinfo {volume} {25}},\ \bibinfo
  {pages} {1157} (\bibinfo {year} {2004})}\BibitemShut {NoStop}%
\bibitem [{\citenamefont {Vanommeslaeghe}\ \emph {et~al.}(2010)\citenamefont
  {Vanommeslaeghe}, \citenamefont {Hatcher}, \citenamefont {Acharya},
  \citenamefont {Kundu}, \citenamefont {Zhong}, \citenamefont {Shim},
  \citenamefont {Darian}, \citenamefont {Guvench}, \citenamefont {Lopes},
  \citenamefont {Vorobyov} \emph {et~al.}}]{vanommeslaeghe2010charmm}%
  \BibitemOpen
  \bibfield  {author} {\bibinfo {author} {\bibfnamefont {K.}~\bibnamefont
  {Vanommeslaeghe}}, \bibinfo {author} {\bibfnamefont {E.}~\bibnamefont
  {Hatcher}}, \bibinfo {author} {\bibfnamefont {C.}~\bibnamefont {Acharya}},
  \bibinfo {author} {\bibfnamefont {S.}~\bibnamefont {Kundu}}, \bibinfo
  {author} {\bibfnamefont {S.}~\bibnamefont {Zhong}}, \bibinfo {author}
  {\bibfnamefont {J.}~\bibnamefont {Shim}}, \bibinfo {author} {\bibfnamefont
  {E.}~\bibnamefont {Darian}}, \bibinfo {author} {\bibfnamefont
  {O.}~\bibnamefont {Guvench}}, \bibinfo {author} {\bibfnamefont
  {P.}~\bibnamefont {Lopes}}, \bibinfo {author} {\bibfnamefont
  {I.}~\bibnamefont {Vorobyov}}, \emph {et~al.},\ }\bibfield  {title} {\bibinfo
  {title} {Charmm general force field: A force field for drug-like molecules
  compatible with the charmm all-atom additive biological force fields},\
  }\href@noop {} {\bibfield  {journal} {\bibinfo  {journal} {Journal of
  computational chemistry}\ }\textbf {\bibinfo {volume} {31}},\ \bibinfo
  {pages} {671} (\bibinfo {year} {2010})}\BibitemShut {NoStop}%
\bibitem [{\citenamefont {Behler}\ and\ \citenamefont
  {Parrinello}(2007)}]{behler2007generalized}%
  \BibitemOpen
  \bibfield  {author} {\bibinfo {author} {\bibfnamefont {J.}~\bibnamefont
  {Behler}}\ and\ \bibinfo {author} {\bibfnamefont {M.}~\bibnamefont
  {Parrinello}},\ }\bibfield  {title} {\bibinfo {title} {Generalized
  neural-network representation of high-dimensional potential-energy
  surfaces},\ }\href@noop {} {\bibfield  {journal} {\bibinfo  {journal}
  {Physical review letters}\ }\textbf {\bibinfo {volume} {98}},\ \bibinfo
  {pages} {146401} (\bibinfo {year} {2007})}\BibitemShut {NoStop}%
\bibitem [{\citenamefont {Musil}\ \emph {et~al.}(2021)\citenamefont {Musil},
  \citenamefont {Grisafi}, \citenamefont {Bart{\'o}k}, \citenamefont {Ortner},
  \citenamefont {Cs{\'a}nyi},\ and\ \citenamefont
  {Ceriotti}}]{musil2021physics}%
  \BibitemOpen
  \bibfield  {author} {\bibinfo {author} {\bibfnamefont {F.}~\bibnamefont
  {Musil}}, \bibinfo {author} {\bibfnamefont {A.}~\bibnamefont {Grisafi}},
  \bibinfo {author} {\bibfnamefont {A.~P.}\ \bibnamefont {Bart{\'o}k}},
  \bibinfo {author} {\bibfnamefont {C.}~\bibnamefont {Ortner}}, \bibinfo
  {author} {\bibfnamefont {G.}~\bibnamefont {Cs{\'a}nyi}},\ and\ \bibinfo
  {author} {\bibfnamefont {M.}~\bibnamefont {Ceriotti}},\ }\bibfield  {title}
  {\bibinfo {title} {Physics-inspired structural representations for molecules
  and materials},\ }\href@noop {} {\bibfield  {journal} {\bibinfo  {journal}
  {arXiv preprint arXiv:2101.04673}\ } (\bibinfo {year} {2021})}\BibitemShut
  {NoStop}%
\bibitem [{\citenamefont {Smith}\ \emph {et~al.}(2017)\citenamefont {Smith},
  \citenamefont {Isayev},\ and\ \citenamefont {Roitberg}}]{smith2017ani}%
  \BibitemOpen
  \bibfield  {author} {\bibinfo {author} {\bibfnamefont {J.~S.}\ \bibnamefont
  {Smith}}, \bibinfo {author} {\bibfnamefont {O.}~\bibnamefont {Isayev}},\ and\
  \bibinfo {author} {\bibfnamefont {A.~E.}\ \bibnamefont {Roitberg}},\
  }\bibfield  {title} {\bibinfo {title} {Ani-1: an extensible neural network
  potential with dft accuracy at force field computational cost},\ }\href@noop
  {} {\bibfield  {journal} {\bibinfo  {journal} {Chemical science}\ }\textbf
  {\bibinfo {volume} {8}},\ \bibinfo {pages} {3192} (\bibinfo {year}
  {2017})}\BibitemShut {NoStop}%
\bibitem [{\citenamefont {Zhang}\ \emph
  {et~al.}(2018{\natexlab{a}})\citenamefont {Zhang}, \citenamefont {Han},
  \citenamefont {Wang}, \citenamefont {Car},\ and\ \citenamefont
  {E}}]{zhang2018deep}%
  \BibitemOpen
  \bibfield  {author} {\bibinfo {author} {\bibfnamefont {L.}~\bibnamefont
  {Zhang}}, \bibinfo {author} {\bibfnamefont {J.}~\bibnamefont {Han}}, \bibinfo
  {author} {\bibfnamefont {H.}~\bibnamefont {Wang}}, \bibinfo {author}
  {\bibfnamefont {R.}~\bibnamefont {Car}},\ and\ \bibinfo {author}
  {\bibfnamefont {W.}~\bibnamefont {E}},\ }\bibfield  {title} {\bibinfo {title}
  {Deep potential molecular dynamics: A scalable model with the accuracy of
  quantum mechanics},\ }\href {https://doi.org/10.1103/PhysRevLett.120.143001}
  {\bibfield  {journal} {\bibinfo  {journal} {Phys. Rev. Lett.}\ }\textbf
  {\bibinfo {volume} {120}},\ \bibinfo {pages} {143001} (\bibinfo {year}
  {2018}{\natexlab{a}})}\BibitemShut {NoStop}%
\bibitem [{\citenamefont {Zhang}\ \emph
  {et~al.}(2018{\natexlab{b}})\citenamefont {Zhang}, \citenamefont {Han},
  \citenamefont {Wang}, \citenamefont {Saidi}, \citenamefont {Car} \emph
  {et~al.}}]{zhang2018end}%
  \BibitemOpen
  \bibfield  {author} {\bibinfo {author} {\bibfnamefont {L.}~\bibnamefont
  {Zhang}}, \bibinfo {author} {\bibfnamefont {J.}~\bibnamefont {Han}}, \bibinfo
  {author} {\bibfnamefont {H.}~\bibnamefont {Wang}}, \bibinfo {author}
  {\bibfnamefont {W.~A.}\ \bibnamefont {Saidi}}, \bibinfo {author}
  {\bibfnamefont {R.}~\bibnamefont {Car}}, \emph {et~al.},\ }\bibfield  {title}
  {\bibinfo {title} {End-to-end symmetry preserving inter-atomic potential
  energy model for finite and extended systems},\ }\href@noop {} {\bibfield
  {journal} {\bibinfo  {journal} {arXiv preprint arXiv:1805.09003}\ } (\bibinfo
  {year} {2018}{\natexlab{b}})}\BibitemShut {NoStop}%
\bibitem [{\citenamefont {Yao}\ \emph {et~al.}(2018)\citenamefont {Yao},
  \citenamefont {Herr}, \citenamefont {Toth}, \citenamefont {Mckintyre},\ and\
  \citenamefont {Parkhill}}]{yao2018tensormol}%
  \BibitemOpen
  \bibfield  {author} {\bibinfo {author} {\bibfnamefont {K.}~\bibnamefont
  {Yao}}, \bibinfo {author} {\bibfnamefont {J.~E.}\ \bibnamefont {Herr}},
  \bibinfo {author} {\bibfnamefont {D.~W.}\ \bibnamefont {Toth}}, \bibinfo
  {author} {\bibfnamefont {R.}~\bibnamefont {Mckintyre}},\ and\ \bibinfo
  {author} {\bibfnamefont {J.}~\bibnamefont {Parkhill}},\ }\bibfield  {title}
  {\bibinfo {title} {The tensormol-0.1 model chemistry: a neural network
  augmented with long-range physics},\ }\href@noop {} {\bibfield  {journal}
  {\bibinfo  {journal} {Chemical science}\ }\textbf {\bibinfo {volume} {9}},\
  \bibinfo {pages} {2261} (\bibinfo {year} {2018})}\BibitemShut {NoStop}%
\bibitem [{\citenamefont {Bart{\'o}k}\ \emph {et~al.}(2010)\citenamefont
  {Bart{\'o}k}, \citenamefont {Payne}, \citenamefont {Kondor},\ and\
  \citenamefont {Cs{\'a}nyi}}]{bartok2010gaussian}%
  \BibitemOpen
  \bibfield  {author} {\bibinfo {author} {\bibfnamefont {A.~P.}\ \bibnamefont
  {Bart{\'o}k}}, \bibinfo {author} {\bibfnamefont {M.~C.}\ \bibnamefont
  {Payne}}, \bibinfo {author} {\bibfnamefont {R.}~\bibnamefont {Kondor}},\ and\
  \bibinfo {author} {\bibfnamefont {G.}~\bibnamefont {Cs{\'a}nyi}},\ }\bibfield
   {title} {\bibinfo {title} {Gaussian approximation potentials: The accuracy
  of quantum mechanics, without the electrons},\ }\href@noop {} {\bibfield
  {journal} {\bibinfo  {journal} {Phys. Rev. Lett.}\ }\textbf {\bibinfo
  {volume} {104}},\ \bibinfo {pages} {136403} (\bibinfo {year}
  {2010})}\BibitemShut {NoStop}%
\bibitem [{\citenamefont {Behler}(2011)}]{behler2011atom}%
  \BibitemOpen
  \bibfield  {author} {\bibinfo {author} {\bibfnamefont {J.}~\bibnamefont
  {Behler}},\ }\bibfield  {title} {\bibinfo {title} {Atom-centered symmetry
  functions for constructing high-dimensional neural network potentials},\
  }\href@noop {} {\bibfield  {journal} {\bibinfo  {journal} {The Journal of
  chemical physics}\ }\textbf {\bibinfo {volume} {134}},\ \bibinfo {pages}
  {074106} (\bibinfo {year} {2011})}\BibitemShut {NoStop}%
\bibitem [{\citenamefont {Yoon}\ \emph {et~al.}(2018)\citenamefont {Yoon},
  \citenamefont {Ha}, \citenamefont {Lee},\ and\ \citenamefont
  {Lee}}]{yoon2018probabilistic}%
  \BibitemOpen
  \bibfield  {author} {\bibinfo {author} {\bibfnamefont {T.~J.}\ \bibnamefont
  {Yoon}}, \bibinfo {author} {\bibfnamefont {M.~Y.}\ \bibnamefont {Ha}},
  \bibinfo {author} {\bibfnamefont {W.~B.}\ \bibnamefont {Lee}},\ and\ \bibinfo
  {author} {\bibfnamefont {Y.-W.}\ \bibnamefont {Lee}},\ }\bibfield  {title}
  {\bibinfo {title} {Probabilistic characterization of the widom delta in
  supercritical region},\ }\href@noop {} {\bibfield  {journal} {\bibinfo
  {journal} {The Journal of chemical physics}\ }\textbf {\bibinfo {volume}
  {149}},\ \bibinfo {pages} {014502} (\bibinfo {year} {2018})}\BibitemShut
  {NoStop}%
\bibitem [{\citenamefont {Schoenholz}\ \emph {et~al.}(2016)\citenamefont
  {Schoenholz}, \citenamefont {Cubuk}, \citenamefont {Sussman}, \citenamefont
  {Kaxiras},\ and\ \citenamefont {Liu}}]{schoenholz2016structural}%
  \BibitemOpen
  \bibfield  {author} {\bibinfo {author} {\bibfnamefont {S.~S.}\ \bibnamefont
  {Schoenholz}}, \bibinfo {author} {\bibfnamefont {E.~D.}\ \bibnamefont
  {Cubuk}}, \bibinfo {author} {\bibfnamefont {D.~M.}\ \bibnamefont {Sussman}},
  \bibinfo {author} {\bibfnamefont {E.}~\bibnamefont {Kaxiras}},\ and\ \bibinfo
  {author} {\bibfnamefont {A.~J.}\ \bibnamefont {Liu}},\ }\bibfield  {title}
  {\bibinfo {title} {A structural approach to relaxation in glassy liquids},\
  }\href@noop {} {\bibfield  {journal} {\bibinfo  {journal} {Nature Physics}\
  }\textbf {\bibinfo {volume} {12}},\ \bibinfo {pages} {469} (\bibinfo {year}
  {2016})}\BibitemShut {NoStop}%
\bibitem [{\citenamefont {Seo}\ \emph {et~al.}(2018)\citenamefont {Seo},
  \citenamefont {Kim}, \citenamefont {Lee}, \citenamefont {Lee},\ and\
  \citenamefont {Lee}}]{seo2018driving}%
  \BibitemOpen
  \bibfield  {author} {\bibinfo {author} {\bibfnamefont {B.}~\bibnamefont
  {Seo}}, \bibinfo {author} {\bibfnamefont {S.}~\bibnamefont {Kim}}, \bibinfo
  {author} {\bibfnamefont {M.}~\bibnamefont {Lee}}, \bibinfo {author}
  {\bibfnamefont {Y.-W.}\ \bibnamefont {Lee}},\ and\ \bibinfo {author}
  {\bibfnamefont {W.~B.}\ \bibnamefont {Lee}},\ }\bibfield  {title} {\bibinfo
  {title} {Driving conformational transitions in the feature space of
  autoencoder neural network},\ }\href@noop {} {\bibfield  {journal} {\bibinfo
  {journal} {The Journal of Physical Chemistry C}\ }\textbf {\bibinfo {volume}
  {122}},\ \bibinfo {pages} {23224} (\bibinfo {year} {2018})}\BibitemShut
  {NoStop}%
\bibitem [{\citenamefont {Boattini}\ \emph {et~al.}(2020)\citenamefont
  {Boattini}, \citenamefont {Mar{\'\i}n-Aguilar}, \citenamefont {Mitra},
  \citenamefont {Foffi}, \citenamefont {Smallenburg},\ and\ \citenamefont
  {Filion}}]{boattini2020autonomously}%
  \BibitemOpen
  \bibfield  {author} {\bibinfo {author} {\bibfnamefont {E.}~\bibnamefont
  {Boattini}}, \bibinfo {author} {\bibfnamefont {S.}~\bibnamefont
  {Mar{\'\i}n-Aguilar}}, \bibinfo {author} {\bibfnamefont {S.}~\bibnamefont
  {Mitra}}, \bibinfo {author} {\bibfnamefont {G.}~\bibnamefont {Foffi}},
  \bibinfo {author} {\bibfnamefont {F.}~\bibnamefont {Smallenburg}},\ and\
  \bibinfo {author} {\bibfnamefont {L.}~\bibnamefont {Filion}},\ }\bibfield
  {title} {\bibinfo {title} {Autonomously revealing hidden local structures in
  supercooled liquids},\ }\href@noop {} {\bibfield  {journal} {\bibinfo
  {journal} {Nature communications}\ }\textbf {\bibinfo {volume} {11}},\
  \bibinfo {pages} {1} (\bibinfo {year} {2020})}\BibitemShut {NoStop}%
\bibitem [{\citenamefont {Gilmer}\ \emph {et~al.}(2017)\citenamefont {Gilmer},
  \citenamefont {Schoenholz}, \citenamefont {Riley}, \citenamefont {Vinyals},\
  and\ \citenamefont {Dahl}}]{gilmer2017neural}%
  \BibitemOpen
  \bibfield  {author} {\bibinfo {author} {\bibfnamefont {J.}~\bibnamefont
  {Gilmer}}, \bibinfo {author} {\bibfnamefont {S.~S.}\ \bibnamefont
  {Schoenholz}}, \bibinfo {author} {\bibfnamefont {P.~F.}\ \bibnamefont
  {Riley}}, \bibinfo {author} {\bibfnamefont {O.}~\bibnamefont {Vinyals}},\
  and\ \bibinfo {author} {\bibfnamefont {G.~E.}\ \bibnamefont {Dahl}},\
  }\bibfield  {title} {\bibinfo {title} {Neural message passing for quantum
  chemistry},\ }in\ \href@noop {} {\emph {\bibinfo {booktitle} {International
  conference on machine learning}}}\ (\bibinfo {organization} {PMLR},\ \bibinfo
  {year} {2017})\ pp.\ \bibinfo {pages} {1263--1272}\BibitemShut {NoStop}%
\bibitem [{\citenamefont {Sch{\"u}tt}\ \emph
  {et~al.}(2017{\natexlab{a}})\citenamefont {Sch{\"u}tt}, \citenamefont
  {Arbabzadah}, \citenamefont {Chmiela}, \citenamefont {M{\"u}ller},\ and\
  \citenamefont {Tkatchenko}}]{schutt2017quantum}%
  \BibitemOpen
  \bibfield  {author} {\bibinfo {author} {\bibfnamefont {K.~T.}\ \bibnamefont
  {Sch{\"u}tt}}, \bibinfo {author} {\bibfnamefont {F.}~\bibnamefont
  {Arbabzadah}}, \bibinfo {author} {\bibfnamefont {S.}~\bibnamefont {Chmiela}},
  \bibinfo {author} {\bibfnamefont {K.~R.}\ \bibnamefont {M{\"u}ller}},\ and\
  \bibinfo {author} {\bibfnamefont {A.}~\bibnamefont {Tkatchenko}},\ }\bibfield
   {title} {\bibinfo {title} {Quantum-chemical insights from deep tensor neural
  networks},\ }\href@noop {} {\bibfield  {journal} {\bibinfo  {journal} {Nature
  communications}\ }\textbf {\bibinfo {volume} {8}},\ \bibinfo {pages} {1}
  (\bibinfo {year} {2017}{\natexlab{a}})}\BibitemShut {NoStop}%
\bibitem [{\citenamefont {Sch{\"u}tt}\ \emph
  {et~al.}(2017{\natexlab{b}})\citenamefont {Sch{\"u}tt}, \citenamefont
  {Kindermans}, \citenamefont {Sauceda}, \citenamefont {Chmiela}, \citenamefont
  {Tkatchenko},\ and\ \citenamefont {M{\"u}ller}}]{schutt2017schnet}%
  \BibitemOpen
  \bibfield  {author} {\bibinfo {author} {\bibfnamefont {K.~T.}\ \bibnamefont
  {Sch{\"u}tt}}, \bibinfo {author} {\bibfnamefont {P.-J.}\ \bibnamefont
  {Kindermans}}, \bibinfo {author} {\bibfnamefont {H.~E.}\ \bibnamefont
  {Sauceda}}, \bibinfo {author} {\bibfnamefont {S.}~\bibnamefont {Chmiela}},
  \bibinfo {author} {\bibfnamefont {A.}~\bibnamefont {Tkatchenko}},\ and\
  \bibinfo {author} {\bibfnamefont {K.-R.}\ \bibnamefont {M{\"u}ller}},\
  }\bibfield  {title} {\bibinfo {title} {Schnet: A continuous-filter
  convolutional neural network for modeling quantum interactions},\ }\href@noop
  {} {\bibfield  {journal} {\bibinfo  {journal} {arXiv preprint
  arXiv:1706.08566}\ } (\bibinfo {year} {2017}{\natexlab{b}})}\BibitemShut
  {NoStop}%
\bibitem [{\citenamefont {Lubbers}\ \emph {et~al.}(2018)\citenamefont
  {Lubbers}, \citenamefont {Smith},\ and\ \citenamefont
  {Barros}}]{lubbers2018hierarchical}%
  \BibitemOpen
  \bibfield  {author} {\bibinfo {author} {\bibfnamefont {N.}~\bibnamefont
  {Lubbers}}, \bibinfo {author} {\bibfnamefont {J.~S.}\ \bibnamefont {Smith}},\
  and\ \bibinfo {author} {\bibfnamefont {K.}~\bibnamefont {Barros}},\
  }\bibfield  {title} {\bibinfo {title} {Hierarchical modeling of molecular
  energies using a deep neural network},\ }\href@noop {} {\bibfield  {journal}
  {\bibinfo  {journal} {The Journal of chemical physics}\ }\textbf {\bibinfo
  {volume} {148}},\ \bibinfo {pages} {241715} (\bibinfo {year}
  {2018})}\BibitemShut {NoStop}%
\bibitem [{\citenamefont {Zubatyuk}\ \emph {et~al.}(2019)\citenamefont
  {Zubatyuk}, \citenamefont {Smith}, \citenamefont {Leszczynski},\ and\
  \citenamefont {Isayev}}]{zubatyuk2019accurate}%
  \BibitemOpen
  \bibfield  {author} {\bibinfo {author} {\bibfnamefont {R.}~\bibnamefont
  {Zubatyuk}}, \bibinfo {author} {\bibfnamefont {J.~S.}\ \bibnamefont {Smith}},
  \bibinfo {author} {\bibfnamefont {J.}~\bibnamefont {Leszczynski}},\ and\
  \bibinfo {author} {\bibfnamefont {O.}~\bibnamefont {Isayev}},\ }\bibfield
  {title} {\bibinfo {title} {Accurate and transferable multitask prediction of
  chemical properties with an atoms-in-molecules neural network},\ }\href@noop
  {} {\bibfield  {journal} {\bibinfo  {journal} {Science advances}\ }\textbf
  {\bibinfo {volume} {5}},\ \bibinfo {pages} {eaav6490} (\bibinfo {year}
  {2019})}\BibitemShut {NoStop}%
\bibitem [{\citenamefont {Bart{\'o}k}\ \emph {et~al.}(2013)\citenamefont
  {Bart{\'o}k}, \citenamefont {Kondor},\ and\ \citenamefont
  {Cs{\'a}nyi}}]{bartok2013representing}%
  \BibitemOpen
  \bibfield  {author} {\bibinfo {author} {\bibfnamefont {A.~P.}\ \bibnamefont
  {Bart{\'o}k}}, \bibinfo {author} {\bibfnamefont {R.}~\bibnamefont {Kondor}},\
  and\ \bibinfo {author} {\bibfnamefont {G.}~\bibnamefont {Cs{\'a}nyi}},\
  }\bibfield  {title} {\bibinfo {title} {On representing chemical
  environments},\ }\href@noop {} {\bibfield  {journal} {\bibinfo  {journal}
  {Physical Review B}\ }\textbf {\bibinfo {volume} {87}},\ \bibinfo {pages}
  {184115} (\bibinfo {year} {2013})}\BibitemShut {NoStop}%
\bibitem [{\citenamefont {Munson}\ and\ \citenamefont
  {Caruana}(2009)}]{munson2009feature}%
  \BibitemOpen
  \bibfield  {author} {\bibinfo {author} {\bibfnamefont {M.~A.}\ \bibnamefont
  {Munson}}\ and\ \bibinfo {author} {\bibfnamefont {R.}~\bibnamefont
  {Caruana}},\ }\bibfield  {title} {\bibinfo {title} {On feature selection,
  bias-variance, and bagging},\ }in\ \href@noop {} {\emph {\bibinfo {booktitle}
  {Joint European Conference on Machine Learning and Knowledge Discovery in
  Databases}}}\ (\bibinfo {organization} {Springer},\ \bibinfo {year} {2009})\
  pp.\ \bibinfo {pages} {144--159}\BibitemShut {NoStop}%
\bibitem [{\citenamefont {Samek}\ and\ \citenamefont
  {M{\"u}ller}(2019)}]{samek2019towards}%
  \BibitemOpen
  \bibfield  {author} {\bibinfo {author} {\bibfnamefont {W.}~\bibnamefont
  {Samek}}\ and\ \bibinfo {author} {\bibfnamefont {K.-R.}\ \bibnamefont
  {M{\"u}ller}},\ }\bibfield  {title} {\bibinfo {title} {Towards explainable
  artificial intelligence},\ }in\ \href@noop {} {\emph {\bibinfo {booktitle}
  {Explainable AI: interpreting, explaining and visualizing deep learning}}}\
  (\bibinfo  {publisher} {Springer},\ \bibinfo {year} {2019})\ pp.\ \bibinfo
  {pages} {5--22}\BibitemShut {NoStop}%
\bibitem [{\citenamefont {Vaswani}\ \emph {et~al.}(2017)\citenamefont
  {Vaswani}, \citenamefont {Shazeer}, \citenamefont {Parmar}, \citenamefont
  {Uszkoreit}, \citenamefont {Jones}, \citenamefont {Gomez}, \citenamefont
  {Kaiser},\ and\ \citenamefont {Polosukhin}}]{vaswani2017attention}%
  \BibitemOpen
  \bibfield  {author} {\bibinfo {author} {\bibfnamefont {A.}~\bibnamefont
  {Vaswani}}, \bibinfo {author} {\bibfnamefont {N.}~\bibnamefont {Shazeer}},
  \bibinfo {author} {\bibfnamefont {N.}~\bibnamefont {Parmar}}, \bibinfo
  {author} {\bibfnamefont {J.}~\bibnamefont {Uszkoreit}}, \bibinfo {author}
  {\bibfnamefont {L.}~\bibnamefont {Jones}}, \bibinfo {author} {\bibfnamefont
  {A.~N.}\ \bibnamefont {Gomez}}, \bibinfo {author} {\bibfnamefont
  {{\L}.}~\bibnamefont {Kaiser}},\ and\ \bibinfo {author} {\bibfnamefont
  {I.}~\bibnamefont {Polosukhin}},\ }\bibfield  {title} {\bibinfo {title}
  {Attention is all you need},\ }in\ \href@noop {} {\emph {\bibinfo {booktitle}
  {Advances in neural information processing systems}}}\ (\bibinfo {year}
  {2017})\ pp.\ \bibinfo {pages} {5998--6008}\BibitemShut {NoStop}%
\bibitem [{\citenamefont {Dosovitskiy}\ \emph {et~al.}(2020)\citenamefont
  {Dosovitskiy}, \citenamefont {Beyer}, \citenamefont {Kolesnikov},
  \citenamefont {Weissenborn}, \citenamefont {Zhai}, \citenamefont
  {Unterthiner}, \citenamefont {Dehghani}, \citenamefont {Minderer},
  \citenamefont {Heigold}, \citenamefont {Gelly} \emph
  {et~al.}}]{dosovitskiy2020image}%
  \BibitemOpen
  \bibfield  {author} {\bibinfo {author} {\bibfnamefont {A.}~\bibnamefont
  {Dosovitskiy}}, \bibinfo {author} {\bibfnamefont {L.}~\bibnamefont {Beyer}},
  \bibinfo {author} {\bibfnamefont {A.}~\bibnamefont {Kolesnikov}}, \bibinfo
  {author} {\bibfnamefont {D.}~\bibnamefont {Weissenborn}}, \bibinfo {author}
  {\bibfnamefont {X.}~\bibnamefont {Zhai}}, \bibinfo {author} {\bibfnamefont
  {T.}~\bibnamefont {Unterthiner}}, \bibinfo {author} {\bibfnamefont
  {M.}~\bibnamefont {Dehghani}}, \bibinfo {author} {\bibfnamefont
  {M.}~\bibnamefont {Minderer}}, \bibinfo {author} {\bibfnamefont
  {G.}~\bibnamefont {Heigold}}, \bibinfo {author} {\bibfnamefont
  {S.}~\bibnamefont {Gelly}}, \emph {et~al.},\ }\bibfield  {title} {\bibinfo
  {title} {An image is worth 16x16 words: Transformers for image recognition at
  scale},\ }\href@noop {} {\bibfield  {journal} {\bibinfo  {journal} {arXiv
  preprint arXiv:2010.11929}\ } (\bibinfo {year} {2020})}\BibitemShut {NoStop}%
\bibitem [{\citenamefont {Liu}\ \emph {et~al.}(2021)\citenamefont {Liu},
  \citenamefont {Lin}, \citenamefont {Cao}, \citenamefont {Hu}, \citenamefont
  {Wei}, \citenamefont {Zhang}, \citenamefont {Lin},\ and\ \citenamefont
  {Guo}}]{liu2021swin}%
  \BibitemOpen
  \bibfield  {author} {\bibinfo {author} {\bibfnamefont {Z.}~\bibnamefont
  {Liu}}, \bibinfo {author} {\bibfnamefont {Y.}~\bibnamefont {Lin}}, \bibinfo
  {author} {\bibfnamefont {Y.}~\bibnamefont {Cao}}, \bibinfo {author}
  {\bibfnamefont {H.}~\bibnamefont {Hu}}, \bibinfo {author} {\bibfnamefont
  {Y.}~\bibnamefont {Wei}}, \bibinfo {author} {\bibfnamefont {Z.}~\bibnamefont
  {Zhang}}, \bibinfo {author} {\bibfnamefont {S.}~\bibnamefont {Lin}},\ and\
  \bibinfo {author} {\bibfnamefont {B.}~\bibnamefont {Guo}},\ }\bibfield
  {title} {\bibinfo {title} {Swin transformer: Hierarchical vision transformer
  using shifted windows},\ }\href@noop {} {\bibfield  {journal} {\bibinfo
  {journal} {arXiv preprint arXiv:2103.14030}\ } (\bibinfo {year}
  {2021})}\BibitemShut {NoStop}%
\bibitem [{\citenamefont {Chen}\ \emph {et~al.}(2021)\citenamefont {Chen},
  \citenamefont {Lu}, \citenamefont {Rajeswaran}, \citenamefont {Lee},
  \citenamefont {Grover}, \citenamefont {Laskin}, \citenamefont {Abbeel},
  \citenamefont {Srinivas},\ and\ \citenamefont {Mordatch}}]{chen2021decision}%
  \BibitemOpen
  \bibfield  {author} {\bibinfo {author} {\bibfnamefont {L.}~\bibnamefont
  {Chen}}, \bibinfo {author} {\bibfnamefont {K.}~\bibnamefont {Lu}}, \bibinfo
  {author} {\bibfnamefont {A.}~\bibnamefont {Rajeswaran}}, \bibinfo {author}
  {\bibfnamefont {K.}~\bibnamefont {Lee}}, \bibinfo {author} {\bibfnamefont
  {A.}~\bibnamefont {Grover}}, \bibinfo {author} {\bibfnamefont
  {M.}~\bibnamefont {Laskin}}, \bibinfo {author} {\bibfnamefont
  {P.}~\bibnamefont {Abbeel}}, \bibinfo {author} {\bibfnamefont
  {A.}~\bibnamefont {Srinivas}},\ and\ \bibinfo {author} {\bibfnamefont
  {I.}~\bibnamefont {Mordatch}},\ }\bibfield  {title} {\bibinfo {title}
  {Decision transformer: Reinforcement learning via sequence modeling},\
  }\href@noop {} {\bibfield  {journal} {\bibinfo  {journal} {arXiv preprint
  arXiv:2106.01345}\ } (\bibinfo {year} {2021})}\BibitemShut {NoStop}%
\bibitem [{\citenamefont {Jumper}\ \emph {et~al.}(2021)\citenamefont {Jumper},
  \citenamefont {Evans}, \citenamefont {Pritzel}, \citenamefont {Green},
  \citenamefont {Figurnov}, \citenamefont {Ronneberger}, \citenamefont
  {Tunyasuvunakool}, \citenamefont {Bates}, \citenamefont {{\v{Z}}{\'\i}dek},
  \citenamefont {Potapenko}, \citenamefont {Bridgland}, \citenamefont {Meyer},
  \citenamefont {Kohl}, \citenamefont {Ballard}, \citenamefont {Cowie},
  \citenamefont {Romera-Paredes}, \citenamefont {Nikolov}, \citenamefont
  {Jain}, \citenamefont {Adler}, \citenamefont {Back}, \citenamefont
  {Petersen}, \citenamefont {Reiman}, \citenamefont {Clancy}, \citenamefont
  {Zielinski}, \citenamefont {Steinegger}, \citenamefont {Pacholska},
  \citenamefont {Berghammer}, \citenamefont {Bodenstein}, \citenamefont
  {Silver}, \citenamefont {Vinyals}, \citenamefont {Senior}, \citenamefont
  {Kavukcuoglu}, \citenamefont {Kohli},\ and\ \citenamefont
  {Hassabis}}]{AlphaFold2021}%
  \BibitemOpen
  \bibfield  {author} {\bibinfo {author} {\bibfnamefont {J.}~\bibnamefont
  {Jumper}}, \bibinfo {author} {\bibfnamefont {R.}~\bibnamefont {Evans}},
  \bibinfo {author} {\bibfnamefont {A.}~\bibnamefont {Pritzel}}, \bibinfo
  {author} {\bibfnamefont {T.}~\bibnamefont {Green}}, \bibinfo {author}
  {\bibfnamefont {M.}~\bibnamefont {Figurnov}}, \bibinfo {author}
  {\bibfnamefont {O.}~\bibnamefont {Ronneberger}}, \bibinfo {author}
  {\bibfnamefont {K.}~\bibnamefont {Tunyasuvunakool}}, \bibinfo {author}
  {\bibfnamefont {R.}~\bibnamefont {Bates}}, \bibinfo {author} {\bibfnamefont
  {A.}~\bibnamefont {{\v{Z}}{\'\i}dek}}, \bibinfo {author} {\bibfnamefont
  {A.}~\bibnamefont {Potapenko}}, \bibinfo {author} {\bibfnamefont
  {A.}~\bibnamefont {Bridgland}}, \bibinfo {author} {\bibfnamefont
  {C.}~\bibnamefont {Meyer}}, \bibinfo {author} {\bibfnamefont {S.~A.~A.}\
  \bibnamefont {Kohl}}, \bibinfo {author} {\bibfnamefont {A.~J.}\ \bibnamefont
  {Ballard}}, \bibinfo {author} {\bibfnamefont {A.}~\bibnamefont {Cowie}},
  \bibinfo {author} {\bibfnamefont {B.}~\bibnamefont {Romera-Paredes}},
  \bibinfo {author} {\bibfnamefont {S.}~\bibnamefont {Nikolov}}, \bibinfo
  {author} {\bibfnamefont {R.}~\bibnamefont {Jain}}, \bibinfo {author}
  {\bibfnamefont {J.}~\bibnamefont {Adler}}, \bibinfo {author} {\bibfnamefont
  {T.}~\bibnamefont {Back}}, \bibinfo {author} {\bibfnamefont {S.}~\bibnamefont
  {Petersen}}, \bibinfo {author} {\bibfnamefont {D.}~\bibnamefont {Reiman}},
  \bibinfo {author} {\bibfnamefont {E.}~\bibnamefont {Clancy}}, \bibinfo
  {author} {\bibfnamefont {M.}~\bibnamefont {Zielinski}}, \bibinfo {author}
  {\bibfnamefont {M.}~\bibnamefont {Steinegger}}, \bibinfo {author}
  {\bibfnamefont {M.}~\bibnamefont {Pacholska}}, \bibinfo {author}
  {\bibfnamefont {T.}~\bibnamefont {Berghammer}}, \bibinfo {author}
  {\bibfnamefont {S.}~\bibnamefont {Bodenstein}}, \bibinfo {author}
  {\bibfnamefont {D.}~\bibnamefont {Silver}}, \bibinfo {author} {\bibfnamefont
  {O.}~\bibnamefont {Vinyals}}, \bibinfo {author} {\bibfnamefont {A.~W.}\
  \bibnamefont {Senior}}, \bibinfo {author} {\bibfnamefont {K.}~\bibnamefont
  {Kavukcuoglu}}, \bibinfo {author} {\bibfnamefont {P.}~\bibnamefont {Kohli}},\
  and\ \bibinfo {author} {\bibfnamefont {D.}~\bibnamefont {Hassabis}},\
  }\bibfield  {title} {\bibinfo {title} {Highly accurate protein structure
  prediction with {AlphaFold}},\ }\bibfield  {journal} {\bibinfo  {journal}
  {Nature}\ }\href {https://doi.org/10.1038/s41586-021-03819-2}
  {10.1038/s41586-021-03819-2} (\bibinfo {year} {2021}),\ \bibinfo {note}
  {(Accelerated article preview)}\BibitemShut {NoStop}%
\bibitem [{\citenamefont {Lee}\ \emph {et~al.}(2019)\citenamefont {Lee},
  \citenamefont {Lee}, \citenamefont {Kim}, \citenamefont {Kosiorek},
  \citenamefont {Choi},\ and\ \citenamefont {Teh}}]{lee2019set}%
  \BibitemOpen
  \bibfield  {author} {\bibinfo {author} {\bibfnamefont {J.}~\bibnamefont
  {Lee}}, \bibinfo {author} {\bibfnamefont {Y.}~\bibnamefont {Lee}}, \bibinfo
  {author} {\bibfnamefont {J.}~\bibnamefont {Kim}}, \bibinfo {author}
  {\bibfnamefont {A.}~\bibnamefont {Kosiorek}}, \bibinfo {author}
  {\bibfnamefont {S.}~\bibnamefont {Choi}},\ and\ \bibinfo {author}
  {\bibfnamefont {Y.~W.}\ \bibnamefont {Teh}},\ }\bibfield  {title} {\bibinfo
  {title} {Set transformer: A framework for attention-based
  permutation-invariant neural networks},\ }in\ \href@noop {} {\emph {\bibinfo
  {booktitle} {International Conference on Machine Learning}}}\ (\bibinfo
  {organization} {PMLR},\ \bibinfo {year} {2019})\ pp.\ \bibinfo {pages}
  {3744--3753}\BibitemShut {NoStop}%
\bibitem [{\citenamefont {Graves}\ \emph {et~al.}(2014)\citenamefont {Graves},
  \citenamefont {Wayne},\ and\ \citenamefont {Danihelka}}]{graves2014neural}%
  \BibitemOpen
  \bibfield  {author} {\bibinfo {author} {\bibfnamefont {A.}~\bibnamefont
  {Graves}}, \bibinfo {author} {\bibfnamefont {G.}~\bibnamefont {Wayne}},\ and\
  \bibinfo {author} {\bibfnamefont {I.}~\bibnamefont {Danihelka}},\ }\bibfield
  {title} {\bibinfo {title} {Neural turing machines},\ }\href@noop {}
  {\bibfield  {journal} {\bibinfo  {journal} {arXiv preprint arXiv:1410.5401}\
  } (\bibinfo {year} {2014})}\BibitemShut {NoStop}%
\bibitem [{\citenamefont {Bahdanau}\ \emph {et~al.}(2014)\citenamefont
  {Bahdanau}, \citenamefont {Cho},\ and\ \citenamefont
  {Bengio}}]{bahdanau2014neural}%
  \BibitemOpen
  \bibfield  {author} {\bibinfo {author} {\bibfnamefont {D.}~\bibnamefont
  {Bahdanau}}, \bibinfo {author} {\bibfnamefont {K.}~\bibnamefont {Cho}},\ and\
  \bibinfo {author} {\bibfnamefont {Y.}~\bibnamefont {Bengio}},\ }\bibfield
  {title} {\bibinfo {title} {Neural machine translation by jointly learning to
  align and translate},\ }\href@noop {} {\bibfield  {journal} {\bibinfo
  {journal} {arXiv preprint arXiv:1409.0473}\ } (\bibinfo {year}
  {2014})}\BibitemShut {NoStop}%
\bibitem [{\citenamefont {Luong}\ \emph {et~al.}(2015)\citenamefont {Luong},
  \citenamefont {Pham},\ and\ \citenamefont {Manning}}]{luong2015effective}%
  \BibitemOpen
  \bibfield  {author} {\bibinfo {author} {\bibfnamefont {M.-T.}\ \bibnamefont
  {Luong}}, \bibinfo {author} {\bibfnamefont {H.}~\bibnamefont {Pham}},\ and\
  \bibinfo {author} {\bibfnamefont {C.~D.}\ \bibnamefont {Manning}},\
  }\bibfield  {title} {\bibinfo {title} {Effective approaches to
  attention-based neural machine translation},\ }\href@noop {} {\bibfield
  {journal} {\bibinfo  {journal} {arXiv preprint arXiv:1508.04025}\ } (\bibinfo
  {year} {2015})}\BibitemShut {NoStop}%
\bibitem [{\citenamefont {Plimpton}(1995)}]{plimpton1995fast}%
  \BibitemOpen
  \bibfield  {author} {\bibinfo {author} {\bibfnamefont {S.}~\bibnamefont
  {Plimpton}},\ }\bibfield  {title} {\bibinfo {title} {Fast parallel algorithms
  for short-range molecular dynamics},\ }\href@noop {} {\bibfield  {journal}
  {\bibinfo  {journal} {Journal of computational physics}\ }\textbf {\bibinfo
  {volume} {117}},\ \bibinfo {pages} {1} (\bibinfo {year} {1995})}\BibitemShut
  {NoStop}%
\bibitem [{\citenamefont {Anderson}\ \emph {et~al.}(2020)\citenamefont
  {Anderson}, \citenamefont {Glaser},\ and\ \citenamefont
  {Glotzer}}]{anderson2020hoomd}%
  \BibitemOpen
  \bibfield  {author} {\bibinfo {author} {\bibfnamefont {J.~A.}\ \bibnamefont
  {Anderson}}, \bibinfo {author} {\bibfnamefont {J.}~\bibnamefont {Glaser}},\
  and\ \bibinfo {author} {\bibfnamefont {S.~C.}\ \bibnamefont {Glotzer}},\
  }\bibfield  {title} {\bibinfo {title} {Hoomd-blue: A python package for
  high-performance molecular dynamics and hard particle monte carlo
  simulations},\ }\href@noop {} {\bibfield  {journal} {\bibinfo  {journal}
  {Computational Materials Science}\ }\textbf {\bibinfo {volume} {173}},\
  \bibinfo {pages} {109363} (\bibinfo {year} {2020})}\BibitemShut {NoStop}%
\bibitem [{\citenamefont {Kingma}\ and\ \citenamefont
  {Ba}(2014)}]{kingma2014adam}%
  \BibitemOpen
  \bibfield  {author} {\bibinfo {author} {\bibfnamefont {D.~P.}\ \bibnamefont
  {Kingma}}\ and\ \bibinfo {author} {\bibfnamefont {J.}~\bibnamefont {Ba}},\
  }\bibfield  {title} {\bibinfo {title} {Adam: A method for stochastic
  optimization},\ }\href@noop {} {\bibfield  {journal} {\bibinfo  {journal}
  {arXiv preprint arXiv:1412.6980}\ } (\bibinfo {year} {2014})}\BibitemShut
  {NoStop}%
\bibitem [{\citenamefont {Groot}\ and\ \citenamefont
  {Warren}(1997)}]{groot1997dissipative}%
  \BibitemOpen
  \bibfield  {author} {\bibinfo {author} {\bibfnamefont {R.~D.}\ \bibnamefont
  {Groot}}\ and\ \bibinfo {author} {\bibfnamefont {P.~B.}\ \bibnamefont
  {Warren}},\ }\bibfield  {title} {\bibinfo {title} {Dissipative particle
  dynamics: Bridging the gap between atomistic and mesoscopic simulation},\
  }\href@noop {} {\bibfield  {journal} {\bibinfo  {journal} {The Journal of
  chemical physics}\ }\textbf {\bibinfo {volume} {107}},\ \bibinfo {pages}
  {4423} (\bibinfo {year} {1997})}\BibitemShut {NoStop}%
\bibitem [{\citenamefont {Stillinger}\ and\ \citenamefont
  {Weber}(1985)}]{stillinger1985computer}%
  \BibitemOpen
  \bibfield  {author} {\bibinfo {author} {\bibfnamefont {F.~H.}\ \bibnamefont
  {Stillinger}}\ and\ \bibinfo {author} {\bibfnamefont {T.~A.}\ \bibnamefont
  {Weber}},\ }\bibfield  {title} {\bibinfo {title} {Computer simulation of
  local order in condensed phases of silicon},\ }\href@noop {} {\bibfield
  {journal} {\bibinfo  {journal} {Physical review B}\ }\textbf {\bibinfo
  {volume} {31}},\ \bibinfo {pages} {5262} (\bibinfo {year}
  {1985})}\BibitemShut {NoStop}%
\bibitem [{\citenamefont {Tersoff}(1988)}]{tersoff1988new}%
  \BibitemOpen
  \bibfield  {author} {\bibinfo {author} {\bibfnamefont {J.}~\bibnamefont
  {Tersoff}},\ }\bibfield  {title} {\bibinfo {title} {New empirical approach
  for the structure and energy of covalent systems},\ }\href@noop {} {\bibfield
   {journal} {\bibinfo  {journal} {Physical review B}\ }\textbf {\bibinfo
  {volume} {37}},\ \bibinfo {pages} {6991} (\bibinfo {year}
  {1988})}\BibitemShut {NoStop}%
\bibitem [{\citenamefont {Bhat}\ \emph {et~al.}(2007)\citenamefont {Bhat},
  \citenamefont {Molinero}, \citenamefont {Soignard}, \citenamefont {Solomon},
  \citenamefont {Sastry}, \citenamefont {Yarger},\ and\ \citenamefont
  {Angell}}]{bhat2007vitrification}%
  \BibitemOpen
  \bibfield  {author} {\bibinfo {author} {\bibfnamefont {M.}~\bibnamefont
  {Bhat}}, \bibinfo {author} {\bibfnamefont {V.}~\bibnamefont {Molinero}},
  \bibinfo {author} {\bibfnamefont {E.}~\bibnamefont {Soignard}}, \bibinfo
  {author} {\bibfnamefont {V.}~\bibnamefont {Solomon}}, \bibinfo {author}
  {\bibfnamefont {S.}~\bibnamefont {Sastry}}, \bibinfo {author} {\bibfnamefont
  {J.}~\bibnamefont {Yarger}},\ and\ \bibinfo {author} {\bibfnamefont
  {C.}~\bibnamefont {Angell}},\ }\bibfield  {title} {\bibinfo {title}
  {Vitrification of a monatomic metallic liquid},\ }\href@noop {} {\bibfield
  {journal} {\bibinfo  {journal} {Nature}\ }\textbf {\bibinfo {volume} {448}},\
  \bibinfo {pages} {787} (\bibinfo {year} {2007})}\BibitemShut {NoStop}%
\bibitem [{\citenamefont {Molinero}\ and\ \citenamefont
  {Moore}(2009)}]{molinero2009water}%
  \BibitemOpen
  \bibfield  {author} {\bibinfo {author} {\bibfnamefont {V.}~\bibnamefont
  {Molinero}}\ and\ \bibinfo {author} {\bibfnamefont {E.~B.}\ \bibnamefont
  {Moore}},\ }\bibfield  {title} {\bibinfo {title} {Water modeled as an
  intermediate element between carbon and silicon},\ }\href@noop {} {\bibfield
  {journal} {\bibinfo  {journal} {The Journal of Physical Chemistry B}\
  }\textbf {\bibinfo {volume} {113}},\ \bibinfo {pages} {4008} (\bibinfo {year}
  {2009})}\BibitemShut {NoStop}%
\bibitem [{\citenamefont {Kob}\ and\ \citenamefont
  {Andersen}(1995)}]{kob1995testing}%
  \BibitemOpen
  \bibfield  {author} {\bibinfo {author} {\bibfnamefont {W.}~\bibnamefont
  {Kob}}\ and\ \bibinfo {author} {\bibfnamefont {H.~C.}\ \bibnamefont
  {Andersen}},\ }\bibfield  {title} {\bibinfo {title} {Testing mode-coupling
  theory for a supercooled binary lennard-jones mixture i: The van hove
  correlation function},\ }\href@noop {} {\bibfield  {journal} {\bibinfo
  {journal} {Physical Review E}\ }\textbf {\bibinfo {volume} {51}},\ \bibinfo
  {pages} {4626} (\bibinfo {year} {1995})}\BibitemShut {NoStop}%
\bibitem [{\citenamefont {Cubuk}\ \emph {et~al.}(2015)\citenamefont {Cubuk},
  \citenamefont {Schoenholz}, \citenamefont {Rieser}, \citenamefont {Malone},
  \citenamefont {Rottler}, \citenamefont {Durian}, \citenamefont {Kaxiras},\
  and\ \citenamefont {Liu}}]{cubuk2015identifying}%
  \BibitemOpen
  \bibfield  {author} {\bibinfo {author} {\bibfnamefont {E.~D.}\ \bibnamefont
  {Cubuk}}, \bibinfo {author} {\bibfnamefont {S.~S.}\ \bibnamefont
  {Schoenholz}}, \bibinfo {author} {\bibfnamefont {J.~M.}\ \bibnamefont
  {Rieser}}, \bibinfo {author} {\bibfnamefont {B.~D.}\ \bibnamefont {Malone}},
  \bibinfo {author} {\bibfnamefont {J.}~\bibnamefont {Rottler}}, \bibinfo
  {author} {\bibfnamefont {D.~J.}\ \bibnamefont {Durian}}, \bibinfo {author}
  {\bibfnamefont {E.}~\bibnamefont {Kaxiras}},\ and\ \bibinfo {author}
  {\bibfnamefont {A.~J.}\ \bibnamefont {Liu}},\ }\bibfield  {title} {\bibinfo
  {title} {Identifying structural flow defects in disordered solids using
  machine-learning methods},\ }\href@noop {} {\bibfield  {journal} {\bibinfo
  {journal} {Physical review letters}\ }\textbf {\bibinfo {volume} {114}},\
  \bibinfo {pages} {108001} (\bibinfo {year} {2015})}\BibitemShut {NoStop}%
\bibitem [{\citenamefont {Schoenholz}\ \emph {et~al.}(2017)\citenamefont
  {Schoenholz}, \citenamefont {Cubuk}, \citenamefont {Kaxiras},\ and\
  \citenamefont {Liu}}]{schoenholz2017relationship}%
  \BibitemOpen
  \bibfield  {author} {\bibinfo {author} {\bibfnamefont {S.~S.}\ \bibnamefont
  {Schoenholz}}, \bibinfo {author} {\bibfnamefont {E.~D.}\ \bibnamefont
  {Cubuk}}, \bibinfo {author} {\bibfnamefont {E.}~\bibnamefont {Kaxiras}},\
  and\ \bibinfo {author} {\bibfnamefont {A.~J.}\ \bibnamefont {Liu}},\
  }\bibfield  {title} {\bibinfo {title} {Relationship between local structure
  and relaxation in out-of-equilibrium glassy systems},\ }\href@noop {}
  {\bibfield  {journal} {\bibinfo  {journal} {Proceedings of the National
  Academy of Sciences}\ }\textbf {\bibinfo {volume} {114}},\ \bibinfo {pages}
  {263} (\bibinfo {year} {2017})}\BibitemShut {NoStop}%
\bibitem [{\citenamefont {Boattini}\ \emph {et~al.}(2019)\citenamefont
  {Boattini}, \citenamefont {Dijkstra},\ and\ \citenamefont
  {Filion}}]{boattini2019unsupervised}%
  \BibitemOpen
  \bibfield  {author} {\bibinfo {author} {\bibfnamefont {E.}~\bibnamefont
  {Boattini}}, \bibinfo {author} {\bibfnamefont {M.}~\bibnamefont {Dijkstra}},\
  and\ \bibinfo {author} {\bibfnamefont {L.}~\bibnamefont {Filion}},\
  }\bibfield  {title} {\bibinfo {title} {Unsupervised learning for local
  structure detection in colloidal systems},\ }\href@noop {} {\bibfield
  {journal} {\bibinfo  {journal} {The Journal of chemical physics}\ }\textbf
  {\bibinfo {volume} {151}},\ \bibinfo {pages} {154901} (\bibinfo {year}
  {2019})}\BibitemShut {NoStop}%
\bibitem [{\citenamefont {Ha}\ \emph {et~al.}(2019)\citenamefont {Ha},
  \citenamefont {Yoon}, \citenamefont {Tlusty}, \citenamefont {Jho},\ and\
  \citenamefont {Lee}}]{ha2019universality}%
  \BibitemOpen
  \bibfield  {author} {\bibinfo {author} {\bibfnamefont {M.~Y.}\ \bibnamefont
  {Ha}}, \bibinfo {author} {\bibfnamefont {T.~J.}\ \bibnamefont {Yoon}},
  \bibinfo {author} {\bibfnamefont {T.}~\bibnamefont {Tlusty}}, \bibinfo
  {author} {\bibfnamefont {Y.}~\bibnamefont {Jho}},\ and\ \bibinfo {author}
  {\bibfnamefont {W.~B.}\ \bibnamefont {Lee}},\ }\bibfield  {title} {\bibinfo
  {title} {Universality, scaling, and collapse in supercritical fluids},\
  }\href@noop {} {\bibfield  {journal} {\bibinfo  {journal} {The journal of
  physical chemistry letters}\ }\textbf {\bibinfo {volume} {11}},\ \bibinfo
  {pages} {451} (\bibinfo {year} {2019})}\BibitemShut {NoStop}%
\end{thebibliography}
%

\end{document}